\documentclass[twocolumn,10pt,showpacs,amsmath,amssymb,aps,prb,superscriptaddress]{revtex4-1}

\usepackage{graphicx}
\usepackage{dcolumn}
\usepackage{bm}

\newcommand{\eg}{\textit{e.g.\ }}

\begin{document}

\title{Momentum distribution and Compton profile by the ab initio GW approximation}

\author{Valerio Olevano}
\affiliation{Institut N\'eel, CNRS \& UJF, Grenoble, France}
\affiliation{European Theoretical Spectroscopy Facility, Grenoble, France}
\author{Andrey Titov}
\affiliation{Institut N\'eel, CNRS \& UJF, Grenoble, France}
\affiliation{L\_Sim, SP2M INAC, CEA, Grenoble, France}
\affiliation{A. M. Prokhorov General Physics Institute, Russian Academy of Sciences, 38 Vavilov street, Moscow 119991 Russia}
\affiliation{European Theoretical Spectroscopy Facility, Grenoble, France}
\author{Massimo Ladisa}
\affiliation{Istituto di Cristallografia, CNR, Bari, Italy}
\affiliation{European Theoretical Spectroscopy Facility, Grenoble, France}
\author{Keijo H\"am\"al\"ainen}
\affiliation{Department of Physics, University of Helsinki, Finland}
\author{Simo Huotari} 
\affiliation{European Synchrotron Radiation Facility, Grenoble, France}
\affiliation{Department of Physics, University of Helsinki, Finland}
\author{Markus Holzmann}
\affiliation{LPTMC, UPMC \& CNRS, Paris, France}
\affiliation{LPMMC, UJF \& CNRS, Grenoble, France}
\affiliation{European Theoretical Spectroscopy Facility, Grenoble, France}

\date{\today}

\begin{abstract}
We present two possible approaches to calculate the momentum distribution $n(p)$ and the Compton profile within the framework of the \textit{ab initio} GW approximation on the self-energy.
The approaches are based on integration of the Green's function along either the real or the imaginary axes.
Examples will be presented on the jellium model and on real bulk sodium.
Advantages and drawbacks of both methods are discussed in comparison with accurate quantum Monte Carlo (QMC) calculations and x-ray Compton scattering experiments.
We illustrate the effect of many-body correlations and disentangle them from band-structure and anisotropy effects by a comparison with density functional theory in the local density approximation.
Our results suggest the use of $G_0W_0$ momentum distributions as reference for future experiments and theory developments.
\end{abstract}

\pacs{71.10.-w, 78.70.Ck, 71.20.Dg, 02.70.Ss}
\keywords{}

\maketitle

\section*{Introduction} 
\begin{figure}
\includegraphics[width=\linewidth]{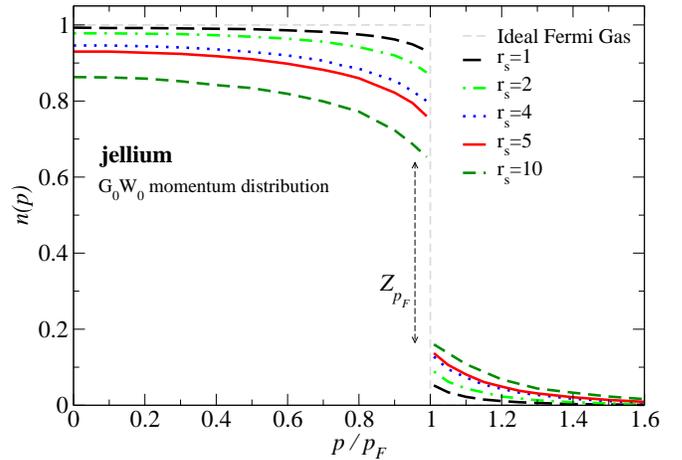}
\caption{The $G_0W_0$ momentum distribution of the jellium model at decreasing densities, as an illustration of increasing correlation effect.
Gray short-dashed line: ideal Fermi gas;
black long-dashed line: $r_s = 1$;
light-green dot-dashed line: $r_s = 2$;
blue dotted line: $r_s = 4$;
red solid line: $r_s = 5$;
dark-green dashed line: $r_s = 10$;
}
\label{nk}
\end{figure}

The momentum distribution $n(p)$, defined as the probability to observe a particle with momentum $p$, is one of the basic quantities in quantum statistical mechanics.
For fermions, it is a quantity showing a direct evidence of the Pauli principle (see Fig.~\ref{nk}).
For an ideal Fermi gas of free non-interacting electrons, the momentum distribution is the Fermi-Dirac distribution.
In equilibrium at zero temperature, it is the step function, i.e., it is 1 for $p$ below the Fermi momentum $p_F$ and 0 above, with a discontinuity equal to one at the Fermi sphere surface. 

When passing from the ideal Fermi gas to a Fermi liquid of interacting fermions,\cite{electrongas} the momentum distribution departs from the perfect step function.
Correlations induce a modification of the distribution with a spill out from lowest to highest momenta, so that the probability to observe an electron at a momentum $p$ larger than the Fermi momentum $p_F$ becomes finite even at zero temperature.
However, $n(p)$ still retains the discontinuity at $p = p_F$, having only its magnitude reduced from 1.

For the homogeneous electron gas (HEG) or jellium, one of the most fundamental models to study electronic correlations, the discontinuity is reduced to the quasiparticle renormalization factor $Z_{p_F}$ calculated at the Fermi momentum.\cite{Luttinger60}
In the high-density limit ($r_s=0$), which is dominated by the kinetic energy, $Z_{p_F}$ approaches the uncorrelated value 1.
The renormalization factor $Z_{p_F}$ is expected to reduce with decreasing density as correlations build up (see Fig.~\ref{nk}).
The discontinuity is still retained at finite densities and vanishes for $r_s \to \infty$.
In particular, for so-called strongly correlated systems, the discontinuity is strongly suppressed.
The modification of the momentum distribution and the reduction of the discontinuity is mainly a correlation effect, unaffected by other, Hartree or exchange, many-body effects.
Thus, the momentum distribution and its discontinuity, unlike other observables such as the band dispersion or the gap, provides a unique and unambiguous quantification of the level of correlation in a system.\cite{hedin65,rice65,pajanne82,lam71,holm98,lantto80,takada91,Ziesche02b,gori-giori02,Holzmannetal,Maebashi11}
Therefore, experimental measurements of $n(p)$ and its discontinuity, are of fundamental importance to test and verify many-body theories.
The level of accuracy in describing correlations by a given theoretical approach can be directly quantified by a comparison of calculated $n(p)$ and $Z$ with experimental results, if they were available.

In this work, we focus on the calculation of the momentum distribution and related quantities within the framework of \textit{ab initio} many-body theory in the GW approximation.
We have studied two possible approaches: the momentum distribution can be calculated by integration of expressions containing the Green's function or the self-energy evaluated on \textit{either} the real \textit{or} the imaginary $\omega$ axis.
Analytically, both integrals provide the same result.
Numerical convergence problems with respect to the integration sampling favour the real-axis integration for evaluating the momentum distribution far away from the Fermi momentum, while the imaginary axis is more accurate near the discontinuity.
Taking bulk sodium and jellium as examples, our results show that the $G_0W_0$ momentum distributions are in good agreement with x-ray Compton scattering experiments\cite{Huotarietal} and also with quantum Monte Carlo results\cite{Holzmannetal}, provided that the appropriate methodology, as analyzed in this article, is used in their calculation. 
As standard $G_0W_0$ are applicable to describe a broad range of realistic systems, results on the momentum distribution and related quantities, \eg the quasiparticle renormalization factor or the Compton profile, can provide accurate reference values and stimulate new experiments.
In turn, since experimental measurements of the Compton profile derivative discontinuity provide an unambiguous quantification of the level of correlation, the limit of validity of the GW approximation can be checked, \eg in strongly-correlated systems, stimulating further theoretical progress.


The momentum distribution $n(p)$ is also the Fourier transform of the \textit{density matrix} $n(r,r')$ which is the fundamental degree of freedom of density-matrix functional theory (DMFT)\cite{Lodwin55,Gilbert75,Ziesche02}.
Since DMFT is an in principle exact theory to calculate the density matrix, it can give access not only to ground-state observables such as Kohn-Sham DFT, but also to some spectral information such as the quasiparticle renormalization factor $Z$ at the Fermi surface, which can be exctracted from the discontinuity of $n(p)$, for example by switching on and off the pseudopotential or the electronic correlations.
Momentum distributions obtained within GW, as presented here, can also represent a good starting point for improving exchange-correlation approximations within DMFT.

In previous works,\cite{Kubo96,Kubo97,Kubo01} Kubo has addressed the momentum distribution in solids by a $G_0W_0$ real axis integration method using the Hamada plasmon-pole model \cite{HamadaHwangFreeman90} which has been criticized\cite{Schuelke97} for not providing an accurate description of the imaginary part of the self-energy, hence of the momentum distribution.
Here we calculate and use the full frequency dependence of the dielectric function.

The paper is organized as follows: 
Section \ref{momentum-distribution} is an introduction to the momentum distribution.
In Sec.~\ref{experiment}, we discuss the relation with the experiment. 
In Sec.~\ref{dft}, we discuss periodic crystalline band-structure and anisotropy effects on $n(p)$. 
In Sec.~\ref{gw}, we consider purely many-body effects and present details of the two $GW$ methodologies used in the calculation.
Finally, we draw conclusions and an outlook.
Unless explicitly specified, we use atomic units (a.u.).

\section{The momentum distribution} 
\label{momentum-distribution}
The spin averaged momentum distribution is defined as
\[
 n(\mathbf{p}) = \frac{1}{2} \sum_\sigma \langle \Psi | \hat{n}_{\mathbf{p}\sigma} | \Psi \rangle
  = \frac{1}{2} \sum_\sigma \langle \Psi | \hat{a}^\dagger_{\mathbf{p}\sigma} \hat{a}_{\mathbf{p}\sigma} | \Psi \rangle
 ,
\]
where $\Psi$ is the ground-state wave function of the system and $\hat{n}_{\mathbf{p}\sigma} = \hat{a}^\dagger_{\mathbf{p}\sigma} \hat{a}_{\mathbf{p}\sigma}$ is the product of creation/annihilation operators for electrons at momentum $\mathbf{p}$ and spin $\sigma$. 
By replacing, in terms of the field operators, $\hat{\psi}_\sigma^\dagger(\mathbf{r}), \hat{\psi}_\sigma(\mathbf{r})$,
\begin{eqnarray}
  n(\mathbf{p}) &=& \frac{1}{2} \sum_\sigma \frac{1}{V} \int d\mathbf{r} d\mathbf{r}' \, e^{-i\mathbf{p}(\mathbf{r} - \mathbf{r}')}
    \langle \Psi | \hat{\psi}_\sigma^\dagger(\mathbf{r}') \hat{\psi}_\sigma(\mathbf{r}) | \Psi \rangle 
  \nonumber \\
  &=& \frac{1}{2V} \int d\mathbf{r} d\mathbf{r}' \, e^{-i\mathbf{p}(\mathbf{r} - \mathbf{r}')} n(\mathbf{r},\mathbf{r}')
  , \label{npnrr}
\end{eqnarray}
the momentum distribution is expressed as the Fourier transform of the reduced single particle density matrix $n(\mathbf{r},\mathbf{r}')$.
One can see that density-matrix functional theory is particularly suitable to calculate the momentum distribution $n(p)$ and its discontinuity $\zeta$, related to the quasiparticle renormalization factor $Z$ at the Fermi surface.

The momentum distribution is the step function only in the \textit{ideal} case of a system of non-interacting free fermions in equilibrium at zero temperature.
The departure from that ideal shape is due to many factors that must be taken into account, namely:
\renewcommand{\labelenumi}{\Alph{enumi})}
\begin{enumerate}
\item finite temperature effects;
\item band-structure effects; 
\item anisotropy of the Fermi surface; 
\item electron-phonon interaction effects;
\item electron-electron correlation many-body effects.
\end{enumerate}
In the next sections we estimate and discuss the weight of each effect compared to the others.

In the paper we use the convention to define the discontinuity of $n(p)$ as $\zeta$, to be distinguished from the quasiparticle renormalization factor $Z$ at the Fermi surface, and affected by all other, \eg band-structure etc., effects.
It should be noted that $\zeta = Z_{p_F}$ only holds in the case of jellium.

\subsection{Non-interacting free fermions and finite-temperature effects}
Consider an unpolarized \textit{non-interacting} and \textit{free} electron gas of given density $n=3/4\pi r_s^3$ and Fermi momentum $p_F = (9\pi/4)^{1/3}/r_s$. 
The electronic structure  is simply provided by perfectly parabolic energy levels, $\varepsilon_\mathbf{p} = p^2 / 2$, and  plane-wave orbitals, $\phi_\mathbf{p}(\mathbf{r}) = e^{i \mathbf{p} \cdot \mathbf{r}} / \sqrt{V}$.
Due to their fermionic nature, electrons obey the Pauli principle, and at equilibrium single-electron states are filled according to the Fermi-Dirac distribution at temperature $T$ and chemical potential $\mu$ with occupation numbers $n(p) = 1 / (e^{(p^2/2 - \mu)/T} +1 )$, which in this case coincides with the momentum distribution. 
In the limit $T \to 0$, $n(p)$ reduces to the step function, $n(p) = \theta(p_F-p)$, with a well-defined jump at $p_F$ of magnitude $\zeta = n(p_F^-) - n(p_F^+)=1$ (gray dashed line in Fig.~\ref{nk}).
Finite temperatures introduce a smoothing of the jump in the region $|p^2-p_F^2|/2m \lesssim T$. 
At room temperature, $|p/p_F-1| \lesssim T/T_F$ where $T_F=p_F^2/2$ is the Fermi temperature.
For metals, we typically have $T/T_F \approx 10^{-2}$ at room temperature, so that temperature effects are fairly negligible for our analysis.

\subsection{Non-interacting electrons in solids: Band-structure effects}
In order to describe  \textit{non-interacting} electrons in a \textit{periodic crystalline solid}, we have to use Bloch energies $\epsilon_{\nu \mathbf{k}}$ and wave functions,
\begin{equation}
\phi_{\nu\mathbf{k}}(\mathbf{r}) = \sum_\mathbf{G} \tilde{\phi}^{\mathbf{G}}_{\nu \mathbf{k}}
e^{i(\mathbf{k}+\mathbf{G})\mathbf{r}},
\label{expansion}
\end{equation} 
where the summation runs over reciprocal lattice vectors $\mathbf{G}$, and $\mathbf{k}$ and $\nu$ are the crystal momentum and the
band index, respectively.

At $T=0$, all states $\{ \nu, \mathbf{k} \}$ are filled up to the Fermi energy $\epsilon_F$, with a step function for the occupation number, $n_{\nu \mathbf{k}} = \theta(\epsilon_F - \epsilon_{\nu \mathbf{k}})$.
As a consequence, the momentum distribution in the independent particle approximation becomes
 \begin{equation}
  n(\mathbf{p}) =  \sum_{\nu \mathbf{k}} \theta(\epsilon_F - \epsilon_{\nu \mathbf{k}}) \sum_\mathbf{G} \delta(\mathbf{p} - \mathbf{k} - \mathbf{G}) \, |\tilde{\phi}^{\mathbf{G}}_{\nu \mathbf{k}}|^2
,
\label{npsolid}
\end{equation}
and will deviate from the ideal-gas step function.

Whereas the Fermi momentum is no longer a well-defined concept, a discontinuity of the momentum distribution still occurs whenever the Bloch energies cross the Fermi surface, defined by $\epsilon_F$. 
Eventually, more than one discontinuity can be present in the momentum distribution when the Fermi surface is organized in several branches.
The contributions of momentum density centered at reciprocal lattice vectors $\mathbf{G} \neq 0$ are called {\it high-momentum (umklapp) components} of the momentum density and can also be seen experimentally in non-jellium-like systems.\cite{huotari2002}
Assuming a partially filled, isotropic valence band, the Fermi sphere of which is entirely contained within the first Brillouin zone (1BZ), \textit{e.g.}, the case of Na, the value of the discontinuity at the Fermi surface is reduced compared to the free-electron gas by the value
\begin{equation}
  \zeta = |\tilde{\phi}^{\mathbf{G}=0}_{\nu=1,\mathbf{k}_F}|^2 \leq 1
\label{zeta}
\end{equation}
With respect to the electron gas, band-structure effects thus reduce the discontinuity to a value provided by the coefficients of the  plane-wave expansion of the orbitals, Eq.~(\ref{expansion}).
The more the system approaches the ideal Fermi gas, the closer the wave functions are to plane waves, and the larger is the discontinuity at the Fermi surface.

Band-structure effects can be calculated by a simple single-electron theory.
We will discuss in detail in Sec.~\ref{dft} how to account them by density functional theory and will provide an illustration of these effects.

\subsection{Non-interacting electrons in solids: Anisotropy effects}
In a real solid, the momentum distribution defined in Eq.~(\ref{npsolid}) is in general anisotropic, and the Fermi surface  is nonspherical.
Whereas angle-resolved momentum distribution measurements \cite{sakurai2011,huotari2009,kontrym2003,kontrym2002,huotari2000} may eventually resolve the full three-dimensional (3D) $n(\mathbf{p})$, Fermi surface anisotropy may prohibit the access to the magnitude $\zeta$ of the $n(p)$ discontinuity by a simple powder-averaged measurement.
We will discuss anisotropy effects also in Sec.~\ref{dft}.

\subsection{Electron-phonon interactions}
An electron in a solid can absorb or emit a phonon with momentum $q$ and in a mode $\mu$ with a probability given by the electron-phonon
coupling $\lambda_{\mu q}$.
The momentum distribution of electrons turns out to be modified as a consequence of such scatterings.
In order to precisely account for such effects, one should evaluate the full electron-phonon scattering matrix and introduce its effect into the momentum distribution as a function of the temperature.
This can be done by constructing an electron-phonon self-energy, for example, in the second Born approximation.
The electron-phonon coupling may lead to a further decrease in the discontinuity of the momentum distribution.
However, since the phonon Debye frequency, $\omega_D,$ is small compared to the Fermi temperature, changes in the momentum distribution due to electron-phonon interactions are expected only within a narrow momentum region $\delta p / p_F \lesssim \omega_D / T_F \approx 10^{-2}$, well beyond the current experimental resolution.
Similar to pure temperature effects on the electronic distribution, we will neglect these effects in the following.

\subsection{Jellium and Fermi liquid behaviour: Electron-electron correlation effects}
\label{correlation}
In an \textit{electron-electron interacting} system, collisions between electrons will in general reduce the discontinuity in the momentum distribution.
This effect can be accounted for only by a theory presenting a good description of correlations and in principle exact to calculate the density matrix $n(r,r')$, or directly the momentum distribution $n(p)$.
Density-matrix functional theory can be such a theory, once a good approximation for the exchange-correlation functional is found.
Density functional theory in the local density approximation (DFT-LDA) or in the generalized gradient approximation (GGA) present a good description of correlations in the electronic density $n(r)$ or other ground-state properties, but Kohn-Sham DFT is not an exact theory to calculate $n(r,r')$ or $n(p)$.

Precise values of the momentum distribution can be obtained from quantum Monte Carlo (QMC) methods.
These methods are based on the full many-body wave functions $\Psi(\mathbf{r_1},\ldots,\mathbf{r_N})$ and correlations are explicitly introduced in many-body Jastrow and backflow potentials.
The reduced single-particle density matrix can be calculated by integrating the many-body wave function,
\[
 n(\mathbf{r},\mathbf{r'}) = N \int d\mathbf{r_2} \ldots d\mathbf{r_N} \, 
   \Psi^*(\mathbf{r},\mathbf{r_2},\ldots,\mathbf{r_N}) \Psi(\mathbf{r'},\mathbf{r_2},\ldots,\mathbf{r_N})
 ,
\]
(with $\int dr_1 \ldots dr_N \Psi^* \Psi = 1$) and the momentum distribution is obtained by Fourier transform, Eq.~(\ref{npnrr}).
Accurate results on the momentum distribution and on the renormalization factor at zero temperature have been obtained by quantum Monte Carlo calculations.\cite{holzmann09,Holzmannetal}
This way is also available for other wave-function-oriented many-body theories, such as, for example, quantum chemistry approaches.

Finally, the momentum distribution can also be calculated in the Green's function approach to many-body theory by
\[
 n(\mathbf{p}) = - \frac{i}{V} \int d\mathbf{r} d\mathbf{r}' \, e^{-i\mathbf{p}(\mathbf{r} - \mathbf{r}')} 
        \int \frac{d\omega}{2\pi} G^<(\mathbf{r},\mathbf{r}',\omega)
 ,
\]
where $G^<$ is the correlation lesser Green's function.
For a system in equilibrium at $T=0$ K, $n(p)$ can be expressed as
\begin{equation}
 n(\mathbf{p}) = \frac{1}{V} \int d\mathbf{r} d\mathbf{r}' \, e^{-i\mathbf{p}(\mathbf{r} - \mathbf{r}')} 
        \int_{-\infty}^\mu d\omega \, A(\mathbf{r},\mathbf{r}',\omega)
 ,\label{npA}
\end{equation}
where $A$ is the spectral function of single-particle excitations that can be obtained from the imaginary part of the Green's function
\begin{equation}
 A(\omega) = - \frac{1}{\pi} \Im G(\omega) \mathop{\mathrm{sgn}}(\omega - \mu)
, \label{AImG}
\end{equation}
and is normalized as follows
\begin{equation}
  \int_{- \infty}^{+ \infty} d\omega \, A(\mathbf{r},\mathbf{r}',\omega) = \delta(\mathbf{r}-\mathbf{r}')
. \label{Asumrule}
\end{equation}
Therefore, $n(\mathbf{p})$ is alternatively given by 
\begin{equation}
 n(\mathbf{p}) = 1 - \frac{1}{V} \int d\mathbf{r} d\mathbf{r}' \, e^{-i\mathbf{p}(\mathbf{r} - \mathbf{r}')} 
        \int_\mu^\infty d\omega \, A(\mathbf{r},\mathbf{r}',\omega)
.
\label{np1-A}
\end{equation}
Thus, the knowledge of the spectral function $A$, or of the Green's function, can provide the momentum distribution upon frequency integration.

The spectral function of a free, non-interacting gas of fermions is diagonal in momentum space, and reduces to a delta function in frequency space
\[
A(\mathbf{p},\omega) = \delta(\omega-p^2/2)
\]
As a consequence, at zero temperature, where the Fermi-Dirac function becomes a step-function, the momentum distribution contains a jump at the Fermi momentum, $p_F$.
The magnitude of the jump is maximal, $\zeta=1$, and directly equal to the strength of the delta-function.

For any approach characterized by an effective, static, single-particle Hamitonian, in particular, the Hartree-Fock method or the Kohn-Sham scheme to DFT, the spectral function is still characterized by exact delta functions in frequency space.
As a consequence, in the homogeneous system, the momentum distribution is still given by the step function of the ideal gas with a jump $\zeta=1$, while in solids the discontinuity is reduced only by other effects, e.g., band-structure or anisotropy.

Correlations will in general lead to a broadening of the single-particle excitation spectra and smooth out the discontinuity in the momentum distribution.
However, for a normal Fermi liquid, the damping of an excitation at the Fermi surface vanishes \cite{Luttinger60} and we approach again a delta-peak in the spectral function, but of reduced weight, the so-called quasiparticle renormalization factor $Z$.
Therefore, we still expect a discontinuity in the momentum distribution, but its magnitude is reduced by a factor $Z < 1$.
In jellium, where band-structure and anisotropy effects are absent, the jump magnitude itself $\zeta$ represents a direct measure of the strength of quasiparticle excitations $Z_{p_F}$ at the Fermi surface.
From Fig.~\ref{nk}, we expect significant changes in the momentum distribution of jellium at metallic densities compared to free electrons, a direct signature of electronic correlations in the system.
In real systems, in particular jellium-like simple metals, the measure of the momentum distribution jump can provide a direct evidence of Fermi liquid behaviour.
For all other metals, the measure of $\zeta$, once disentangled all other, e.g., band-structure, etc., reduction effects, can provide an estimate of $Z_{k_F}$ and thus a quantification of correlation effects and possible deviation from Fermi liquid behaviour.

In Sec.~\ref{gw}, we will enter into the details of a momentum distribution calculation in the framework of many-body perturbation theory, including correlations by the GW approximation.

\begin{figure}
\includegraphics[width=\linewidth]{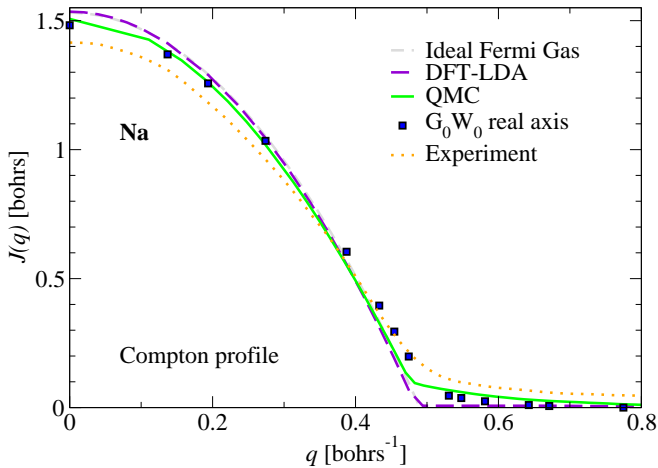}
\caption{
Compton profile of valence electrons in sodium.
Orange dotted line: experiment;
gray short-dashed line: ideal Fermi gas;
green line: QMC calculation;
violet dashed line: DFT-LDA calculation;
blue squares: $G_0W_0$ real-axis calculation.
The ideal Fermi gas result is nearly indistinguishable from the DFT-LDA line.
}
\label{cp}
\end{figure}

\section{The Compton profile and experimental measurements}
\label{experiment}


An interesting technique that can provide an experimental indirect access to the momentum distribution is Compton scattering of x-rays,\cite{cooperbook} a branch of non-resonant inelastic x-ray scattering spectroscopy (IXSS) performed with hard x rays (with energies 10-100 keV) in the limit of large transferred momentum and energy.
IXSS measures the probability for inelastic x-ray scattering, quantified by the double-differential scattering cross section $d^2 \sigma /d \Omega d\omega$ in the solid angle element $d \Omega$ and transferred energy interval $d \omega$. 
At values of $\omega$ much larger than the electron binding energy, the experiment is in the regime of the so-called \textit{impulse approximation} \cite{eisenberger1970}, where the differential scattering cross-section is proportional to the \textit{Compton profile} $J(\mathbf{q})$,
\begin{equation}
 \frac{d^2 \sigma}{d \Omega d\omega} \propto J(\mathbf{q})
,
\end{equation}
where $\mathbf{q}$ is the electron momentum vector that points in the direction of the scattering vector.
The Compton profile is directly related to the momentum distribution and defined as the projection of $n(\mathbf{p})$ onto the direction $\hat{\mathbf{q}}$ of the scattering vector,
\[
 J(\mathbf{q}) = \frac{2}{\bar{n}} \int \frac{d\mathbf{p}}{(2\pi)^3} \, \delta(q - \mathbf{p} \cdot \hat{\mathbf{q}}) n(\mathbf{p})
,
\]
where $\bar{n} = 2 \int d\mathbf{p}/(2 \pi)^3 \, n(\mathbf{p})$ is the average electron density (factor 2 is for spin).

Substantial simplification is obtained assuming an isotropic system $n(\mathbf{p}) = n(p)$.
In this case, we have
\begin{equation}
 J(q) = \frac{1}{2 \pi^2 \bar{n}} \int_{q}^\infty  dp \, p \, n(p)
\label{jqaveraged}
\end{equation}
and $n(p)$ is determined by simple differentiation of the Compton profile
\[
\label{equ:np-from-CP}
n(p) = -\frac{2 \pi^2 \bar{n}}{p} \left. \frac{dJ(q)}{dq} \right|_{q=p}.
\]
Metals with highly isotropic momentum distribution are therefore most suited for measurements of the momentum distribution via Compton scattering.
Notice that we are using the convention by which the Compton profile is normalized to 1,
\[
 \int dq \, J(q) = 1
 .
\]

For example, in the case of the ideal Fermi gas the momentum distribution $n(p)$ is the step function $n(p) = \theta(p_F-p)$ with a discontinuity $\zeta=1$ at the Fermi momentum $p_F$.
The associated Compton profile is an inverted parabola for $q < p_F$,
\[
  J(q) = \frac{3}{4 p_F^3} \left(p_F^2 - q^2\right) \, \theta(p_F-q) 
 ,
\]
and vanishes for $q>p_F$ (see Fig.~\ref{cp}).
It presents a discontinuity of the first derivative at the Fermi momentum $q=p_F$, which is related to the discontinuity $\zeta$ of the momentum distribution $n(p)$.
A direct measure of the discontinuity of the first derivative of the Compton profile $dJ/dq$ provides the discontinuity of the momentum distribution which we are interested in.

Measurements of the Compton profile of solid sodium have been presented in  Ref.~\onlinecite{Huotarietal}.
These experiments were performed at the beamline ID16 of the European Synchrotron Radiation Facility.
Details of the beamline, the spectrometer, sample preparation, and data analysis are given in Refs.~\onlinecite{verbeni2009,Huotarietal,huotari2005}.
To get Compton profiles of valence electrons, the core-electrons contribution must be subtracted from the experimental signal.
Core-electron IXSS spectra were calculated by a quasi-self-consistent field (QSCF) approximation \cite{issolah1988} and the real-space-multiple scattering approach {\tt FEFFq} \cite{soininen2005}, both found in agreement with each other and the measured Compton spectra.\cite{Huotarietal}
In Fig.~\ref{cp} we present the resulting experimental Compton profile of sodium valence electrons together with various theoretical results.
With respect to the Compton profile of an ideal Fermi gas, we observe a departure from the perfect inverted parabola and a reduction of the discontinuity in the first derivative at the Fermi momentum. 
In the next sections we will analyse in detail the influence of band structure and correlations on the Compton profile and the underlying momentum distribution within the GW approach.

\begin{figure}
\includegraphics[width=0.5\linewidth]{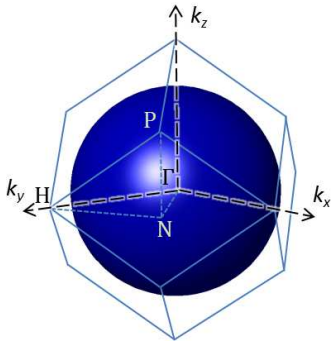}
\caption{The Fermi surface in sodium, obtained from a Kohn-Sham DFT-LDA calculation.}
\label{fermisurface}
\end{figure}

\section{Band-structure and anisotropy effects by DFT-LDA}
\label{dft}

Density-functional theory is an in principle exact theory to calculate the ground-state electron density $n(\mathbf{r})=\langle \Psi_0 |
\hat{\psi}^\dag(\mathbf{r}) \hat{\psi}(\mathbf{r}) | \Psi_0 \rangle$ together with the ground-state energy.
However, neither the off-diagonal elements of the density matrix $n(\mathbf{r},\mathbf{r'})=\langle \Psi_0 | \hat{\psi}^\dag(\mathbf{r}') \hat{\psi}(\mathbf{r}) | \Psi_0 \rangle$, nor its Fourier transform, the momentum distribution, can in general be described within DFT, similar to quasiparticle properties.

Nevertheless, the Kohn-Sham scheme of DFT can be used to evaluate the leading-order band-structure and anisotropy effects to the momentum distribution.
This can be done by replacing Kohn-Sham wave functions and energies into Eq.~(\ref{npsolid}) where correlation effects are neglected.
Whenever the Kohn-Sham wave functions are close to the exact quasiparticle ones, and quasiparticle corrections do not significantly modify the Kohn-Sham Fermi surface, Eq.~(\ref{npsolid}) is expected to provide a good description of band-structure and anisotropy effects, as well as the position of the discontinuity.
This is the case in alkali metals and simple semiconductors.\cite{HybertsenLouie}
A breakdown of this picture was found in solids which contain shallow $d$ electrons\cite{GattiBrunevalOlevano}, and may
also be expected for systems with shallow $f$ electrons.
In these cases, the Kohn-Sham wave functions of $d$ electrons are in general not close to quasiparticle ones.
Further, a quasiparticle shift of $d$ levels may significantly modify the DFT Kohn-Sham Fermi surface, and so the position of the discontinuity.

The system studied here, sodium, is significantly simpler, and  DFT provides a reasonable estimate of band-structure and anisotropy effects.
We run a standard Martin-Trouillers pseudopotential DFT-LDA calculation on a plane-waves basis set, with a cutoff of 24 Ryd on the kinetic energy, fixing the lattice parameter to the experimental value 8.108 a.u.\ (corresponding to an average density of $r_s=3.99$).
Using a Monkhorst-Pack $8 \times 8 \times 8$ grid of $k$ points to represent the Brillouin zone, the electron density is converged  at self-consistency.
Larger sets of $k$ points were used to calculate the band plot, the Fermi surface, and the momentum distribution.

\begin{figure}
\includegraphics[width=\linewidth]{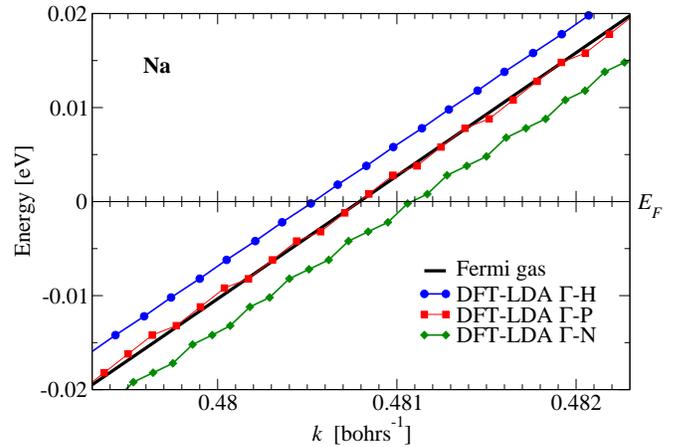}
\caption{Anisotropy of the Fermi surface. Sodium first band energy vs. $k$:
Blue circles and line: along the $\Gamma$-H direction; green squares and line: along the $\Gamma$-P direction; red diamonds and line: along the $\Gamma$-N direction. Black line: Fermi gas at $r_s = 3.9917$ energy levels.}
\label{kf}
\end{figure}

We start with an analysis of the anisotropy effects, as provided by DFT.
The Fermi surface (Fig.~\ref{fermisurface}) is very close to a perfect sphere with a deviation of 0.2\% only.
This result shows a close similarity between sodium and the jellium model.
In order to estimate the maximum anisotropy of the Fermi surface in sodium, we calculated the Kohn-Sham energies on a very fine k-mesh across the Fermi level and along the three high symmetry directions, $\Gamma$-H, $\Gamma$-P and $\Gamma$-N (Fig.~\ref{kf}).  The anisotropy at the Fermi surface is evaluated to be $\Delta k=5.5 \cdot 10^{-4}$ Bohr$^{-1}$.
Such a value is well within the accuracy of the IXSS experiment. 
Therefore, anisotropy effects in sodium are currently not detectable\cite{anisotropy}, and we do not expect more insights from
angle resolved measurements on single crystals compared to much simpler powder averaged measurements.

\begin{figure}
\includegraphics[width=\linewidth]{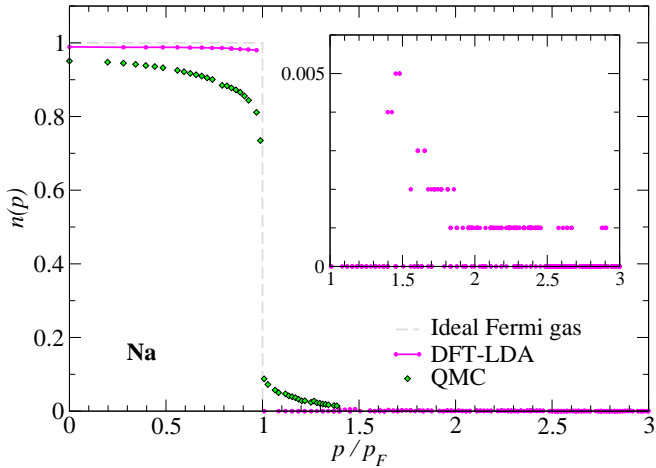}
\caption{The momentum distribution of Na determined by a DFT-LDA calculation, as compared to the ideal Fermi gas step function and to the QMC calculation.
Data have been rounded to $10^{-3}$ corresponding to the accuracy of the calculation.
}
\label{nanpdft}
\end{figure}

\begin{figure}
\includegraphics[width=\linewidth]{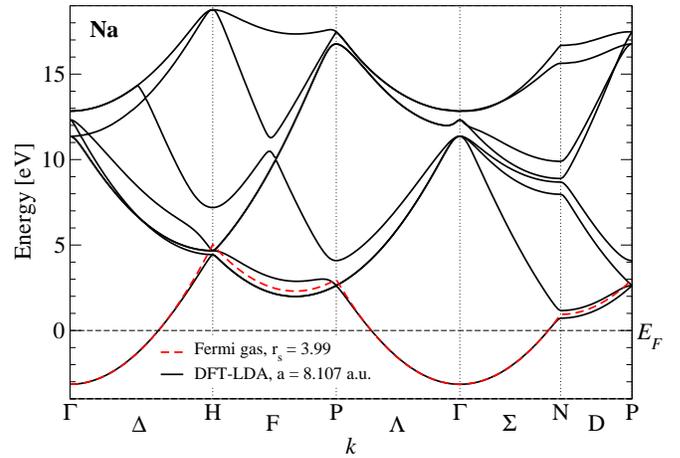} 
\caption{Band structure of sodium. Black line: DFT-LDA Kohn-Sham band structure (the experimental lattice parameter at room temperature was used). Red dashed line: dispersion of the ideal Fermi gas evaluated at the corresponding density, $r_s=3.99$.}
\label{bandplot}
\end{figure}

The momentum distribution calculated using Eq.~(\ref{npsolid}) and Kohn-Sham DFT-LDA wave functions is shown in Fig.~\ref{nanpdft}.
A very small deviation from the momentum distribution of the ideal Fermi gas is observed below and above $p_F$.
The DFT-LDA discontinuity is evaluated to be $\zeta^{\text{Na}}_{\text{DFT}}=0.98(1)$, slightly below the ideal Fermi gas value.
An analysis of the Kohn-Sham wave functions shows that the plane wave coefficients $\tilde{\phi}^{\mathbf{G}=0}_{\nu=1,\mathbf{k}}$ in sodium are 0.99 for all k-points around $\mathbf{k}_F$.
This value remains almost the same for states at the bottom of the first band and up to 1~eV above the Fermi level.
Therefore, wave functions of the first band are very close to plane waves in a wide range of $k$, consistent with  the band-structure (Fig.~\ref{bandplot}).
The first band in sodium is an almost perfect parabola which superimposes with the dispersion curve of the ideal Fermi gas.
A difference between the two curves is appreciable only above the Fermi level and close to the BZ boundary.
The DFT band-width is only 0.04 eV larger in the ideal Fermi gas than in sodium, a value below the standard accuracy of a pseudopotential
calculation.

Beyond the first Brillouin zone, band-structure effects introduce deviations from the momentum distribution of an ideal Fermi gas which exactly vanishes above $p_F$ (Fig.~\ref{nanpdft} inset).
This is typical for crystalline solids and  arises  from  the Fourier components of the Bloch wave functions with $\mathbf{G} \ne 0$.
The weights of these components (smaller than 0.005 in sodium) quantify the  deviations of  the crystalline wave function from a perfect plane wave $e^{-i\mathbf{k} \cdot \mathbf{r}}/\sqrt{V}$.  

We can conclude that both band-structure and anisotropy effects on the momentum distribution are very small in sodium.
They affect $n(p)$ by less than 0.02, with the maximum value achieved at the level of the discontinuity.
The DFT momentum distribution of sodium remains close to that of the ideal Fermi gas, significantly different from  QMC results\cite{Huotarietal} (Fig.~\ref{nanpdft}).
From Eq.~(\ref{jqaveraged}), we  obtain the directional averaged Compton profile.
Quantitative comparison with the measured Compton profile confirms that the bare DFT does not provide accurate momentum distributions or Compton profiles (Fig.~(\ref{cp})).

\section{Correlations effects by the GW approximation}
\label{gw}

The Green's function $G$ or the spectral function $A$ entering into Eq.~(\ref{npA}) for $n(p)$, can be calculated in the Hedin's GW approximation\cite{Hedin} where the self-energy is approximated by
\begin{equation}
 \Sigma^\mathrm{GW}(\mathbf{r},\mathbf{r}',\omega) = i\int \frac{d\omega'}{2\pi} \, G(\mathbf{r},\mathbf{r}',\omega-\omega') W(\mathbf{r},\mathbf{r}',\omega')
, \label{SGWrrpomega}
\end{equation}
in terms of the Green's function $G$ and the dynamically screened interaction $W(\omega) = \varepsilon^{-1}(\omega) v$.
Here, $v$ is the bare Coulomb potential and $\varepsilon^{-1}(\omega)$ the inverse of the dielectric function taken in the random phase approximation (RPA).\cite{Lindhard}
In an iterative scheme, starting with a trial Green's function, $G^0$, for example, obtained from the noninteracting band structure or from DFT Kohn-Sham orbitals, one calculates the RPA polarizability $\Pi^\mathrm{RPA} = G^0G^0$, followed by the dielectric function $\varepsilon^\mathrm{RPA} = 1 - v \Pi^\mathrm{RPA}$, and the RPA dynamically screened interaction $W^\mathrm{RPA} = \varepsilon_\mathrm{RPA}^{-1}(\omega) v$.
With these ingredients, the GW self-energy for the first iteration  can be calculated according to Eq.~(\ref{SGWrrpomega}), and the corresponding Green's function is determined by solving  Dyson's equation,
\begin{equation}
  G= G^{0} + G^{0} \Sigma G
. \label{dyson}
\end{equation}
In the fully self-consistent GW approximation, further iterations up to self-consistency in the self-energy and Green's function
must be performed.
If applied to realistic electronic systems, fully self-consistency is in general a too difficult task, since all functions depend on all space-time coordinates.

Within the so-called $G_0W_0$ approximation, observables are calculated from the self-energy and Green's function of the first iteration.
The $G_0W_0$ approximation turned out\cite{AulburJonsson} to be a useful approach for calculating quasiparticle electronic structure of realistic systems, yielding  band-gaps in solids in very good overall agreement with the experiment.
Different schemes have been explored to obtain partial self-consistency\cite{Kotani,Bruneval}, but fully self-consistent solutions are only reported for jellium.\cite{HolmVonBarth}
Self-consistent calculations are appealing since they eliminate the dependence of the results on the initial trial of the Green's function.
However, the issue  whether a full self-consistent GW really improve upon $G_0W_0$, remains still controversial, and likely depends on the observable one is interested in.

\begin{figure}
\includegraphics[width=\linewidth]{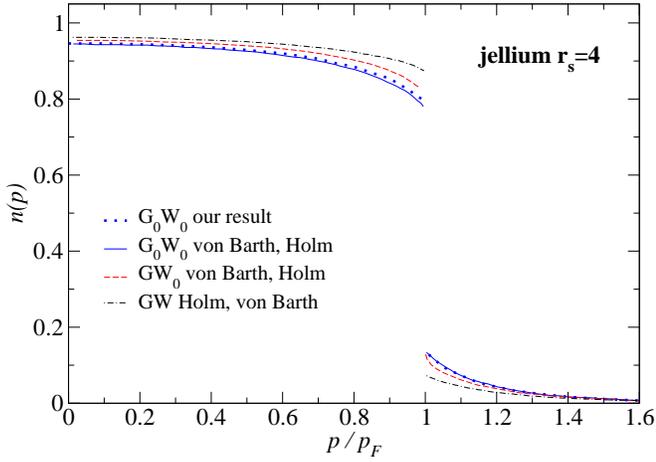}
\caption{Comparison of our momentum distribution with previous results on jellium at $r_s=4$.
Blue dots: our $G_0W_0$ result;
blue solid line: $G_0W_0$ result of von Barth and Holm (Ref.~\onlinecite{VonBarthHolm});
red dashed line: $GW^0$ result of von Barth and Holm (Ref.~\onlinecite{VonBarthHolm});
black dot-dashed line: $GW$ result of Holm and von Barth (Ref.~\onlinecite{HolmVonBarth}).
}
\label{nkrs4}
\end{figure}

\subsection{GW $n(p)$ in jellium as compared to QMC}
The first $G_0W_0$ electronic structure calculation on the jellium model was carried out by Hedin\cite{Hedin}.
Lundqvist\cite{Lundqvist1967,Lundqvist1968} addressed also the momentum distribution.
More recent calculations, involving self-consistency issues within a Gaussian basis set, were done by von Barth and Holm.\cite{VonBarthHolm,HolmVonBarth}
Here, we calculated the $G_0W_0$ momentum distribution following the original approach by Hedin.\cite{Hedin}

In Fig.~\ref{nk}, we show our calculated momentum distributions for jellium for a large range of densities.
Starting from lowest $r_s=1$, passing by metallic densities $r_s=$2-5, up to $r_s=10$ with gradually increasing correlation, we
observe the departure of the momentum distribution from the ideal Fermi gas step function and an increasing spill out to higher momenta, associated with a reduction of the discontinuity, the quasiparticle renormalization factor $Z$ at the Fermi momentum.
We found $Z_{p_F} = 0.86$ for $r_s=1$, as in Ref.~\onlinecite{Hedin}, down to $Z_{p_F} = 0.45$ for $r_s=10$.

In Fig.~\ref{nkrs4}, we compare our results to those of Refs.~\onlinecite{VonBarthHolm,HolmVonBarth}.
Our $G_0W_0$ $n(p)$ is overall in good agreement with the $G_0W_0$ result by Ref.~\onlinecite{VonBarthHolm}.
The small discrepancy at values of $p \lesssim p_F$ is not due to the different scheme (Gaussian basis set versus sampling of $\omega$ axis), but rather to the fact that for $p<p_F$ we used Eq.~(\ref{np1-A}) instead of Eq.~(\ref{npA}),\footnote{
It should be noticed that Eq.~(\ref{np1-A}) is equivalent to Eq.~(\ref{npA}) only in the exact theory or for number-conserving approximations. 
Non-self-consistent $G_0W_0$ is not conserving.
However, the difference one can find in $G_0W_0$ between the two formulas cannot be larger than 0.001 at metallic densities, as it has been found by Schindlmayr \textit{et al.}\cite{Schindlmayr01}
This value is not appreciable here due to our numerical error which is not smaller than 0.01 both in the real sodium and in the jellium calculations.
}
a point that will be clarified later for the more critical case of sodium.

\begin{figure}
\includegraphics[width=\linewidth]{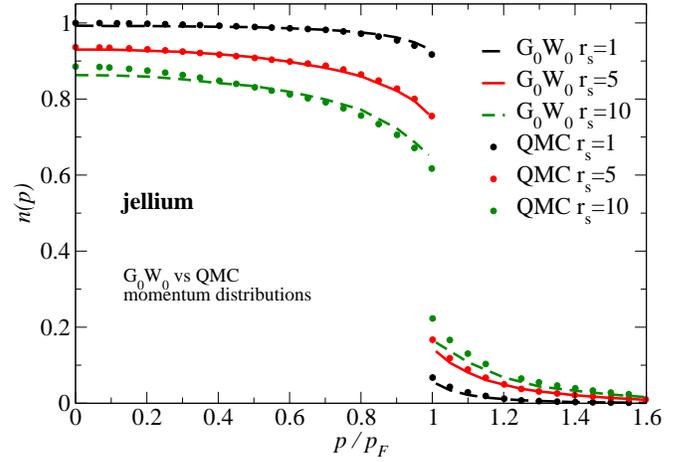}
\caption{Comparison of the $G_0W_0$ and QMC (Ref.~\onlinecite{Holzmannetal}) momentum distribution on jellium at several densities.
Black long-dashed line: $r_s = 1$;
red solid line: $r_s = 5$;
dark-green dashed line: $r_s = 10$.
}
\label{nkgwqmc}
\end{figure}

We then notice that any higher level of self-consistency, both on $G$ and $W$, reduces the level of correlation, and thus increases the
discontinuity and $Z_{p_F}$.
From comparison with quantum Monte Carlo results of Ref.~\onlinecite{Holzmannetal} (Fig.~\ref{nkgwqmc}), we observe that the $G_0W_0$ result is already in surprisingly good agreement with QMC, especially for metallic densities.
Even for the most correlated density, $r_s = 10$, the QMC discontinuity $Z=0.40$ is not far from our $G_0W_0$ prediction of $0.45$.
On the other hand, any improvement towards a fully self-consistent GW has lead to an increase of the discontinuity in disagreement with QMC.
Notice that Refs.~\onlinecite{VonBarthHolm,HolmVonBarth} provide fully self-consistent solutions with respect to all
degrees of freedom.

\subsection{GW $n(p)$ in solids}
For $G_0W_0$ non self-consistent calculations on realistic solids, it is convenient\cite{HybertsenLouie} to start from a Kohn-Sham DFT-LDA eigenvalues $\epsilon^\mathrm{KS}_{\nu\mathbf{k}}$ and eigenfunctions $\phi^\mathrm{KS}_{\nu\mathbf{k}}(\mathbf{r})$ zero-order electronic structure
\begin{equation}
  G^{0}(\mathbf{r},\mathbf{r}',\omega) = \sum_{\nu\mathbf{k}} 
   \frac{\phi_{\nu\mathbf{k}}(\mathbf{r}) \phi^*_{\nu\mathbf{k}}(\mathbf{r}')}
   {\omega - \epsilon_{\nu \mathbf{k}} + i \eta \mathop{\mathrm{sgn}}(\epsilon_{\nu \mathbf{k}} - \mu)}
,
\end{equation}
used to obtain the screened interaction and the GW self-energy, as described above.
However, in order to avoid double counting of exchange-correlation effects already taken into account by the DFT exchange-correlation potential $v^{xc}$, the Dyson equation to calculate the Green's function $G$ in first iteration read as
\begin{equation}
  G= G^{0} + G^{0} \left( \Sigma - v^{xc} \right) G
, \label{dysonvxc}
\end{equation}

In a solid, spectral functions, as well as all other many-body quantities, e.g.\ $G$ and $\Sigma$, can be conveniently described in terms of an orthonormal set of Bloch wave functions, $\phi_{\nu\mathbf{k}}(\mathbf{r})=\sum_\mathbf{G} \tilde{\phi}^{\mathbf{G}}_{\nu \mathbf{k}} e^{i(\mathbf{k}+\mathbf{G})\mathbf{r}}$, as already provided by the Kohn-Sham eigenfunctions.
Expressing the spectral function in this basis set,
\begin{equation}
  A(\mathbf{r},\mathbf{r}',\omega) = \sum_{\nu \mathbf{k} \nu' \mathbf{k}'} \phi_{\nu\mathbf{k}}(\mathbf{r}) \phi^*_{\nu'\mathbf{k}'}(\mathbf{r}') A_{\nu\nu'}(\mathbf{k},\mathbf{k}',\omega)
,
\label{Aband}
\end{equation}
the momentum distribution read as
\begin{equation}
 n(\mathbf{p}) = \sum_{\nu\nu'\mathbf{k}\mathbf{G}} \delta(\mathbf{p} - \mathbf{k} - \mathbf{G})
   \tilde{\phi}^{\mathbf{G}}_{\nu \mathbf{k}} \tilde{\phi}^{\mathbf{G}*}_{\nu' \mathbf{k}}
   \int_{-\infty}^\mu d\omega \, A_{\nu\nu'}(\mathbf{k},\mathbf{k},\omega)
.
\end{equation}
In sodium and other nearly free-electron systems, the self-energy operator, as well as the spectral function $A$, are almost diagonal on $\nu$ and $\nu'$, and the neglection of non-diagonal elements is justified\cite{LundqvistLyden71} in the expression for the momentum distribution,
\begin{eqnarray}
  n(\mathbf{p}) &=& \sum_{\nu\mathbf{k}\mathbf{G}} \delta(\mathbf{p} - \mathbf{k} - \mathbf{G}) \,
   |\tilde{\phi}^{\mathbf{G}}_{\nu \mathbf{k}}|^2 \, n_{\nu\mathbf{k}} 
   , \label{np} \\
   n_{\nu\mathbf{k}} &=& \int_{-\infty}^\mu d\omega \, A_{\nu}(\mathbf{k},\omega)
  , \label{nkrealaxis}
\end{eqnarray}
where $A_{\nu}(\mathbf{k},\omega) \equiv  A_{\nu\nu}(\mathbf{k},\mathbf{k},\omega)$.
This approximation does not hold in systems where some of the true quasiparticle wave functions differ from the DFT wave functions, and  the former expression involving also non-diagonal elements, $A_{\nu\nu'}(\mathbf{k},\mathbf{k},\omega)$, should be used.

In Eqs.~(\ref{np}) and (\ref{nkrealaxis}), $n_{\nu\mathbf{k}}$  provides the \textit{correlation} contribution to the momentum distribution from each band.
For any uncorrelated system, $n_{\nu\mathbf{k}}=1$ for all the states within the Fermi surface, $\epsilon_{\nu\mathbf{k}} < \mu$, and 0 elsewhere (step function) and the strength of the quasiparticle excitation remains $Z=1$ as for free fermions.
However, band-structure effects reduce the discontinuity in the momentum distribution, $\zeta<Z$, already for non-interacting electrons, Eq.~(\ref{zeta}).
Therefore, it is important to distinguish $\zeta$ [the jump in $n(\mathbf{p})$], from the quasiparticle renormalization factor $Z$ (the jump in  $n_{\nu\mathbf{k}}$) quantifying the strength of excitations at the Fermi surface.
Whereas both coincide for a homogeneous system, e.g., jellium, they are different in real solids.
Within leading order, correlations induce deviations of $n_{\nu\mathbf{k}}$ from the step function, whereas band-structure effects are already contained in the weights $\tilde{\phi}^{\mathbf{G}}_{\nu \mathbf{k}}$ inside the summation Eq.~(\ref{np}) over all bands, reciprocal lattice, and BZ vectors.

\subsection{GW $n(p)$ by integration on the real $\omega$-axis}

\begin{figure}
\includegraphics[width=\linewidth]{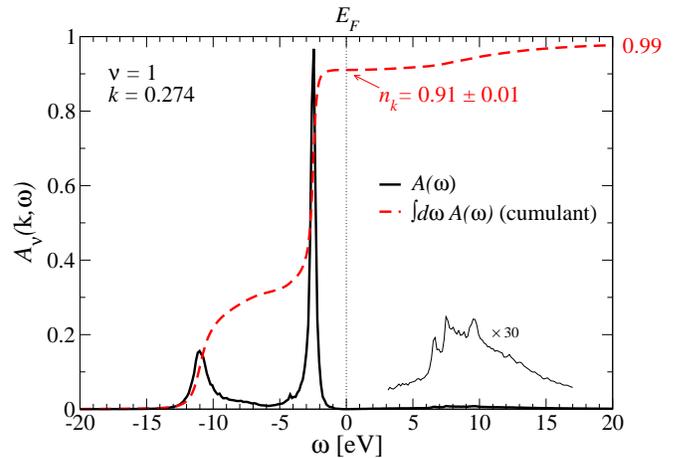}
\caption{Spectral function $A_{\nu}(k,\omega)$ of sodium for the first band ($\nu=1$) and $k=0.274$ (solid black line);
Cumulative sum of the spectral function, $\int_{-\infty}^\omega d\omega' \, A(\omega')$ (red dashed line).
The crossing of the cumulant to the Fermi level provides the value of $n_k$.
The total integral of the spectral function, which should be 1, provides the error on $n_k$.
A zoom of the $A_{\nu}(k,\omega)$ within the energy interval 3--17~eV is shown by a thin black line.
}
\label{sf}
\end{figure}

Neglecting off-diagonal contributions in the band index, the spectral function, Eq.~(\ref{Aband}), can be expressed as
\begin{equation}
 A_\nu(\mathbf{k},\omega) = \frac{1}{\pi} \frac{|\Im \Sigma_\nu(\mathbf{k},\omega)|}
{(\omega - \epsilon_{\nu \mathbf{k}} + v^{xc}_{\nu\mathbf{k}} - \Re \Sigma_\nu(\mathbf{k},\omega))^2 + (\Im \Sigma_\nu(\mathbf{k},\omega))^2}
\label{Aknu}
\end{equation}
in terms of the band-diagonal self-energy, $\Sigma_\nu(\mathbf{k},\omega) \equiv \Sigma_{\nu\nu}(\mathbf{k},\omega)$, calculated within $G_0W_0$, the Kohn-Sham eigenvalues $\epsilon_{\nu \mathbf{k}}$, and the matrix elements of the DFT-LDA exchange-correlation potential
\begin{equation}
 v^{xc}_{\nu\mathbf{k}} = \langle \phi_{\nu\mathbf{k}} | v^{xc} | \phi_{\nu\mathbf{k}}  \rangle.
\end{equation}
From the spectral function, we can calculate $n_{\nu \mathbf{k}}$ by integration along the real axis, Eq.~(\ref{nkrealaxis}). Finally, the momentum distribution is evaluated from Eq.~(\ref{np}) using the plane wave coefficients $\tilde{\phi}^{\mathbf{G}}_{\nu \mathbf{k}}$ of the underlying DFT-LDA calculation.
For the $G_0W_0$ calculation we have used 50 bands, a cutoff of 5 H (both on the wave functions and on the dimension of the polarizability matrices).
The dielectric function has been calculated on 800 frequencies along the real $\omega$ axis, up to 2 Ha, and 10 Gauss-Legendre knots frequencies along the imaginary axis.
With these parameters, $n(p)$ at a given $p$ is converged within $10^{-2}$.

In Fig.~\ref{sf} we show the $G_0W_0$ spectral function $A_{\nu}(\mathbf{k},\omega)$ for $k/k_F \simeq 0.57$. It contains a quasiparticle peak (the most intense feature) and some satellites.
The distribution of the spectral weight among the various structures can be read off by the integrated spectral function (its cumulant).
In this case, the quasiparticle peak has a weight $Z \simeq 0.6$, low- and high-energy satellites have weights of 0.3 and of 0.1, respectively.
According to the sum rule Eq.~(\ref{Asumrule}), the total weight of the spectral function $A_{\nu}(\mathbf{k},\omega)$ is 1.
The difference between the integrated spectral weight and unity is the numerical error of our calculation $\approx 0.01$ in this case.
The value of $n_{\nu \mathbf{k}}$ is graphically provided by the crossing of the cumulant with the Fermi energy, i.e., $n_{\nu \mathbf{k}} = 0.91 \pm 0.01$ for $k$=0.274 and $\nu$=1 (Fig.~\ref{sf}).

\begin{figure}
\includegraphics[width=\linewidth]{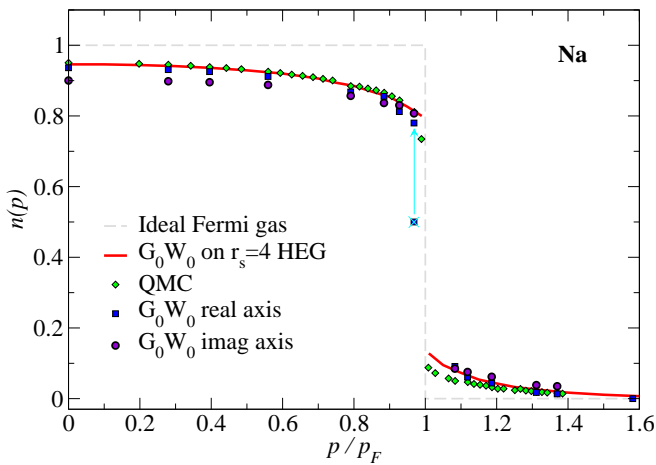}
\caption{The $G_0W_0$ momentum distribution in Na calculated along the real (blue squares) and imaginary (violet circles) axes, as compared to the quantum Monte Carlo (green diamonds) data (Ref.~\onlinecite{Huotarietal}). In the case of integration along the real axis and for $k=0.475$ we show both the direct result of the integration [$n(p) = 0.5$] and the corrected value of $n(p)=0.78$ (the correction procedure is described in the text). 
We show also the momentum distribution in the ideal Fermi gas (gray dashed line).
}
\label{nanpgw}
\end{figure}

The momentum distribution calculated by the $G_0W_0$ method using the real axis integration is shown in Fig.~\ref{nanpgw} at the example of sodium.  
The results are in good agreement with the QMC calculation over a large range of $p$.
When looking at the way $n(p)$ is calculated in a many-body approach, Eq.~(\ref{nkrealaxis}), the good agreement with QMC implies that the $G_0W_0$ approximation reproduces the correct spectral weight repartition between the quasiparticle peak and the rest (satellites).
It also implies that the energy position of the quasiparticle peak with respect to the Fermi energy is correctly reproduced by $G_0W_0$.
However, \textit{it does not yet imply a correct energy position of satellites}.
Since $n(p)$ is just only sensitive to the satellites' spectral weight, GW turns out to be a good approximation to describe the momentum distribution, regardless its ability to correctly describe the satellites' energy position.

\begin{figure}
\includegraphics[width=\linewidth]{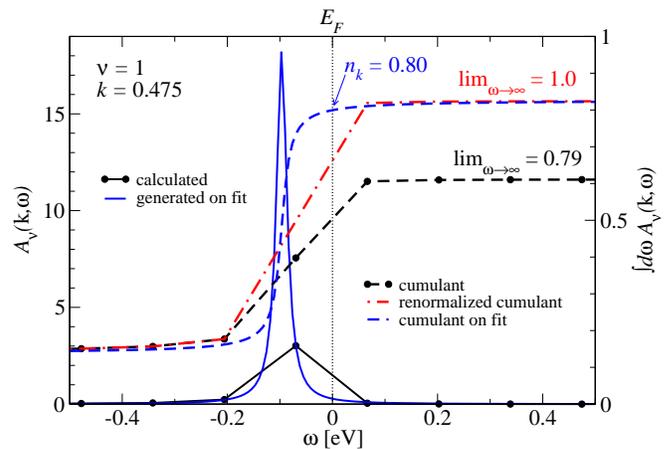}
\caption{The spectral function for $k=0.475$ (close to $k_F$) and $\nu=1$. Black continuous line and circles: $G_0W_0$ spectral function calculated on a coarse $\omega$ mesh; black dashed line and circles: integrated spectral weight of the calculated $G_0W_0$ spectral function, $\int_{-\infty}^\omega d\omega' \, A(\omega')$; red dot-dashed line: renormalized integrated spectral weight; blue continous curve: QP peak generated by a fit over a Lorentzian; blue dashed line: cumulative sum over the QP fit generated curve.}
\label{sf-disc}
\end{figure}

Notice that close to the Fermi surface, the straightforward integration over real frequencies using a fixed discretization grid fails, e.g., the point at $n(p=0.475)=0.5$  in Fig.~\ref{nanpgw} obtained from direct integration provides a too low value for the momentum occupation. 
The observed underestimation of $n_{\nu \mathbf{k}}$ near the Fermi surface is an artifact of the coarse sampling along the $\omega$ axis.
Since the quasiparticle lifetime tends to infinity at the Fermi surface, the width of the quasi particle peak becomes increasingly narrow, and is described by less and less points.
As we can see from  Fig.~\ref{sf-disc}, the quasiparticle peak for $k=0.475$  is actually described by only three points on the $\omega$-mesh underlying our calculation. 
As a consequence, the integral $\int d\omega A(\omega)$ (black dashed line and circles in Fig.~\ref{sf-disc}) is highly inaccurate.
The normalization of $A(\omega)$ turns out to be 0.79 instead of 1, the remaining spectral weight, 0.21, is lost due to the undersampling.
Using Eq.~(\ref{np1-A}) which involves only the spectral function of unoccupied states offers certain improvement.
However, one should keep in mind that the finite $\omega$-mesh employed in calculations on realistic systems might not accurately capture all features of the spectral function for energies and momenta very close to the Fermi surface.  

The drawback related to the coarse $\omega$-mesh can be corrected by a more appropriated evaluation, without the need of recalculation of the spectral function on a more expensive fine $\omega$-mesh.
Based on the assumption that, for points $k$ close to the Fermi surface, the (negligible) spectral weight of satellites is correctly reproduced by our coarse grid, we expand the self-energy inside the spectral function, Eq.~(\ref{Aknu}), around their values at the grid-points.
This justifies the use of a Lorentzian form for the spectral function around the quasiparticle peak.
The spectral weight is then essentially lost by the poor description of the quasiparticle peak on the finite $\omega$ mesh.
This procedure corrects the values of momentum distribution close to the Fermi surface, in particular, the value at $n(p=0.475)=0.78$ is significantly increased compared with the value of the direct integration (see Fig.~\ref{nanpgw}).

In Fig.~\ref{sf-disc},  we illustrate this procedure where a Lorentzian form for the quasiparticle peak is assumed
\begin{equation}
 l(\omega) = Z_\mathrm{QP} \frac{1}{\pi} \frac{\eta}{(\omega-\omega_0)^2 + \eta^2}
 ,
\end{equation}
and the parameters $Z_\mathrm{QP}$, $\omega_0$, and $\eta$ are determined  from a fit of the $G_0W_0$ data such to obtain a total weight of 1 to satisfy Eq.~(\ref{Asumrule}).
The final result of this procedure is the shift of the $n(p)$ point evidenced by the blue arrow in Fig.~\ref{sf-disc}.
The weight above the Fermi level of the fitted QP peak (blue line in Fig.~\ref{sf-disc}) is reduced, as compared to the weight under the the black $G_0W_0$ points connected by direct lines.
The obtained integrated spectral weight, shown by the blue dashed line in Fig.~\ref{sf-disc}, provides a value of $n_{\nu k}=0.80$ for $\nu=1$ and $k=0.475$.
The final value of $n(p)= 0.78$ is then obtained as the product of $n_{\nu k}$ and of the wave function coefficient.

\begin{figure}
\includegraphics[width=\linewidth]{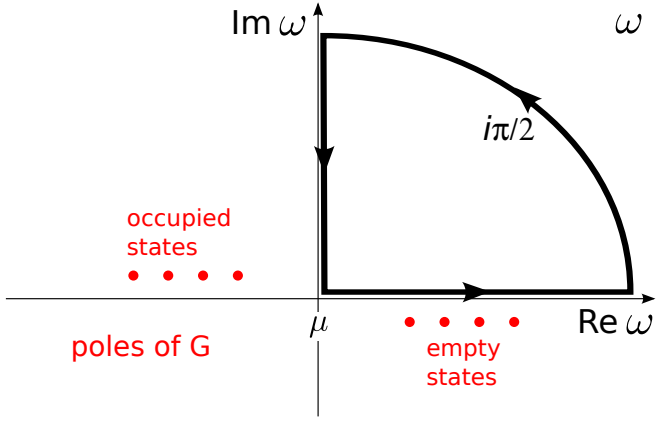}
\caption{Integration contour: the integration along the real axis is replaced by the integration along the imaginary axis and a quarter circle.}
\label{contour}
\end{figure}

\subsection{GW $n(p)$ by integration on the imaginary $\omega$-axis}

A different strategy to obtain the momentum distribution consists in deforming the frequency integration in the complex plane toward an imaginary $\omega$-axis integration.
Therefore, using the normalization of $A$ Eq.~(\ref{Asumrule}), we write $n_{\nu \mathbf{k}}$ in the following form
\begin{equation}
  n_{\nu \mathbf{k}} = 1 - \int_{\mu}^{+ \infty} d\omega \, A_{\nu}(\mathbf{k},\omega) 
  = 1 + \frac{1}{\pi} \int_{\mu}^{+ \infty} d\omega \,  \Im  G_{\nu}(\mathbf{k},\omega)
.
\end{equation}
Since the Green's function $G_{\nu}(\mathbf{k},\omega)$ is analytic in the upper part of the complex plane for $\omega>\mu$, the frequency integral along the real axis can be now deformed to the positive imaginary axis (Fig.~\ref{contour}).
Picking up a contribution $i \pi /2$ from the quarter circle at infinite distance, we obtain\cite{VonBarthHolm}
\begin{equation}
  n_{\nu \mathbf{k}} = \frac{1}{2} + \frac{1}{\pi} \int_{0}^{+ \infty} d\omega \, \Re  G_{\nu}(\mathbf{k},\mu + i\omega) 
. \label{nkimag}
\end{equation}

Thus we can replace the integral along the real $\omega$-axis of the imaginary part of $G$ by an integral along the imaginary $\omega$-axis of the real part of $G$, which is in general a rather smooth function.
The real part of the Green's function $\Re G$ can again be evaluated from $\Sigma$ solving the Dyson equation (\ref{dysonvxc}),
\begin{equation}
 \Re G_\nu(\mathbf{k},\mu + i \omega) = \frac{\mu - \epsilon_{\nu \mathbf{k}} + v^{xc}_{\nu\mathbf{k}} - \Re \Sigma_\nu(\mathbf{k},\mu + i \omega)}{\left( \mu - \epsilon_{\nu \mathbf{k}} + v^{xc}_{\nu\mathbf{k}} - \Re \Sigma \right)^2 + \left( \omega - \Im \Sigma \right)^2 }
, \label{ReG}
\end{equation}
In this approach, also the GW self-energy is calculated by imaginary frequency integration in Eq.~(\ref{SGWrrpomega}).
The computational cost is  significantly reduced with respect to the integration on the real axis.

\begin{figure}
\includegraphics[width=\linewidth]{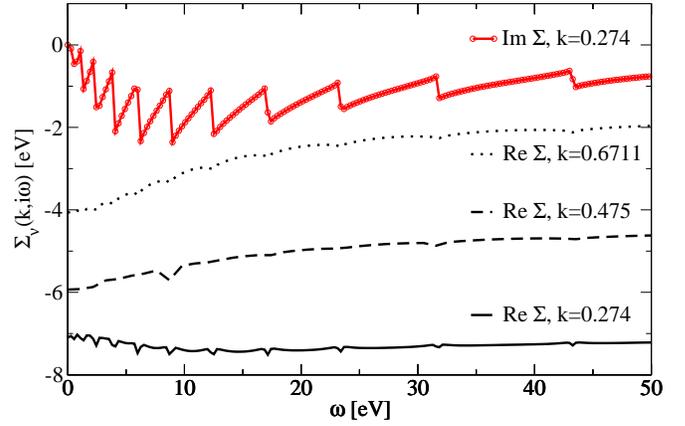}
\caption{Real and imaginary parts of the self-energy along the $\omega$ frequency imaginary axis calculated by the non-self-consistent $G_0W_0$ approach for sodium.
We show the real parts for three k-points: $k=0.274$ (black continuous line); $k=0.475$ (black dashed line); $k=0.6711$ (blue dotted line).
The imaginary part is shown only for $k=0.274$ (red continuous line and circles).
\label{sigmaiomega}
}
\end{figure} 

The calculated real and imaginary parts of the self-energy $\Sigma(\omega)$ along the imaginary frequency $\omega$-axis are shown in Fig.~\ref{sigmaiomega} at the example of sodium.
One can see that the real part of the self-energy is a much smoother quantity on the imaginary axis, in contrast to the imaginary part\cite{dynamicaleffects}.
Close to the Fermi surface, where the imaginary part vanishes, imaginary-axis integration does not suffer from the problems encountered in the integration along the real axis.
So the description of the discontinuity in $n(p)$ is better in the imaginary- than in the real-axis integration (Fig.~\ref{nanpgw}).
However, undersampling artifacts may still occur in the description of states far away from the Fermi surface.
Here, in contrast to the real frequency integration, we have not corrected these artifacts, which explains the differences between the two integration methods in Fig.~\ref{nanpgw}, with the real axis integration more in agreement with the QMC momentum distribution at $p$ far away from the Fermi surface.


\subsection{Contributions from higher bands induced by GW correlations}
\begin{figure}
\includegraphics[width=\linewidth]{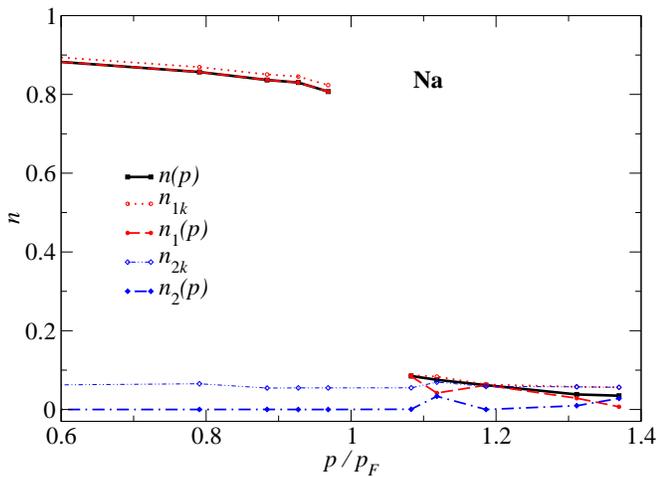}
\caption{The momentum distribution $n(p)$ of sodium determined by a $G_0W_0$ calculation with integration along the imaginary axis (black squares and solid line). 
Contribution $n_1(p)$ from the first (red filled circles and dashed line) and its distribution $n_{1k}$ (red empty circles and dotted line), and from the second band [$n_1(p)$, blue filled diamonds and dashed line] together with the $n_{2k}$ (blue empty diamonds and dotted line) distribution. Lines are guide to the eyes.
\label{nanpgwimag}
}
\end{figure}

In Fig.~\ref{nanpgwimag} we analyze the contributions of the different bands to the momentum distribution.
We show the $n_{\nu \mathbf{k}}$ curves, as calculated from Eq.~(\ref{nkimag}), for the first $\nu=1$ band (dotted red lines) and for the second $\nu=2$ band (blue double-dot-dashed line), the $n_\nu(p)$ curves including band-structure effects, Eq.~(\ref{np}), for the first (red dashed line) and second (blue dot-dashed line) bands, together with the full $n(p)$ (black solid line).
At the non-interacting level in sodium only the first band contributes to $n(p)$ [see Eq.~(\ref{npsolid}), Fig.~\ref{bandplot} and \ref{nanpdft}].
Correlation effects may induce contributions from higher bands.
From Fig.~\ref{nanpgwimag} we observe that in sodium the first $\nu=1$ band provides the dominant contribution inside the first Brillouin zone, in particular for $p \lesssim  p_F$.  
Correlation effects induce a contribution $n_{2k}$ from the second band of an almost constant 0.08.
However band-structure effects, taken into account by the $|\tilde{\phi}^{\mathbf{G}=0}_{\nu=2,\mathbf{k}}|^2$ coefficients in Eq.~(\ref{np}), are more intense than in the first band (which is almost parabolic and planewaves like), and  depress this contribution to negligible values.
The contribution of the second band starts to be at the level of the first only around $p=1.1 \, p_F$ and for $p \gtrsim 1.3 \, p_F$ where the $|\tilde{\phi}^{\mathbf{G=0}}_{\nu=2,\mathbf{k}}|^{2}$ coefficients are significant.
Higher bands ($\nu>2$) provide a negligible contribution. 

The discontinuity $\zeta$ in sodium is dominated by the first band only.
The magnitude of  the discontinuity is therefore given by
\begin{equation}
  \zeta = |\tilde{\phi}^{\mathbf{G=0}}_{\nu=1, \mathbf{k}_F}|^{2} Z_{\nu=1,\mathbf{k}_F},
\label{zetaZ}
\end{equation}
where $ Z_{\nu=1,\mathbf{k}_F}$ is the renormalization factor of the first band quasiparticle peak at the Fermi surface.
Since $|\tilde{\phi}^{\mathbf{G=0}}_{\nu=1, \mathbf{k}_F}|^{2} \simeq 0.99$, band-structure effects introduce only a small 1\% reduction of the discontinuity $\zeta$ in the momentum distribution of sodium.

\subsection{Anisotropies in the GW renormalization factor}

In general in solids the renormalization factor is a function of the wave-vector $\mathbf{k}$ direction.
In Table~\ref{zanisotropy}, we have calculated the $G_0W_0$ $Z_{\mathbf{k}_F}$ for the first sodium band,
\begin{equation}
  Z_{\mathbf{k}_{F}}= \left( 1 - \left. \frac{\partial \Re \Sigma(\mathbf{k}_F,\omega)}{\partial \omega} \right|_{\omega=\mu}  \right)^{-1}
,
\label{zkf}
\end{equation}
along the main symmetry directions in the BZ.
In sodium anisotropy effects are small also on many-body quantities, leading to variations in $Z$ of less than 1 \%.
The $G_0W_0$ value of $Z=0.65 \pm 0.01$ in sodium is directly comparable with the $G_0W_0$ value of $Z = 0.64$ in jellium at the sodium average density of $r_s \simeq 4$.
Valence electrons of sodium look very much like a good realization in nature of the ideal homogeneous electron model, both from the point of view of non-interacting, as well as many-body interacting observables.
We can then compare the $G_0W_0$ value to many different theoretical prediction\cite{hedin65,rice65,pajanne82,lam71,holm98,lantto80,takada91,Ziesche02b,gori-giori02,Maebashi11}, including the latest QMC value for jellium\cite{Holzmannetal}, $0.64 \pm 0.01$, which is more accurate than that reported for sodium\cite{Huotarietal} ($0.70 \pm 0.02$).
The experimental value $0.58 \pm 0.07$ from Ref.~\onlinecite{Huotarietal}, clearly eliminates several important approximations, such as fully self-consistent GW\cite{holm98} as well as the so-called on-shell RPA approximation.\cite{pajanne82,lam71}

\begin{table}[t]
\begin{tabular}{ll}
\hline \hline
$k$-direction & $Z^\textrm{Na}_{\mathbf{k}_F}$ ($G_0W_0$) \\
\hline
$\Gamma$-N & 0.649(1) \\
$\Gamma$-P & 0.643(1) \\
$\Gamma$-H & 0.649(1) \\
\hline \hline
\end{tabular}
\caption{Anisotropy of the $G_0W_0$ quasiparticle renormalization factor in sodium.}
\label{zanisotropy}
\end{table}

\subsection{GW Compton profile in sodium}

In Fig.~\ref{cp} we present the GW Compton profile as compared to the ideal Fermi gas, DFT-LDA, QMC, and the IXSS experiment.
Both the ideal Fermi gas and the DFT-LDA Compton profile appear like an inverted parabola, with a discontinuity in the first derivative equal to 1 (or nearly for DFT-LDA) at the Fermi momentum.
The DFT-LDA Compton profile is almost coincident with the ideal Fermi gas curve, so that band-structure effects are even less appreciable than in the momentum distribution: small differences in the two $n(p)$ are further smoothed after the integration to get at $J(q)$.

The discontinuity is clearly reduced in the QMC sodium pseudopotential and in $G_0W_0$ calculations.
Both QMC and $G_0W_0$ discontinuities are in agreement with the experimental discontinuity within its error bar, mostly dictated by the experimental momentum resolution and statistical accuracy.
The $G_0W_0$ real axis integration Compton profile is practically coincident with the Slater-Jastrow QMC result, except at $p_F$ where it provides a lower discontinuity.
A more accurate backflow QMC calculation in sodium would probably further reduce the discontinuity, like it is the case in jellium.
As to the comparison with the experiment:
in principle, most possible systematic errors that may influence the experimental Compton profile, such as the finite experimental $q$-resolution, tend to reduce the experimental value of $J(0)$.
For this reason, the experimental Compton profile should be rather regarded as the lowest bottom extremum for small $q$ and a highest limiting value for large $q$.


\section*{Conclusions and outlook}

We have presented two possible ways within the \textit{ab initio} $G_0W_0$ approximation to calculate the momentum distribution $n(p)$, the Compton profile $J(q)$ and their discontinuities associated to the quasiparticle renormalization factor $Z_{p_F}$.
We have analyzed and discussed the advantages and drawbacks of both approaches in comparison with QMC calculations and x-ray Compton scattering measurements on bulk sodium.
All the analyzed quantities have been found in good agreement with both QMC and the experiment.

In sodium, we have found that $n(p)$ and $J(q)$ are mostly determined by the first band and are very weakly affected by band-structure and anisotropy effects, in contrast to other alkalines, \eg Li\cite{Filippi99,Liexp1,liexp2}.
This confirms  that the valence electrons of sodium, in ambient conditions, almost perfectly realize the jellium model, even considering its electron dynamics\cite{huotari2011,cazzaniga2011}.
Since jellium is one of the most fundamental models to study electronic correlations, a reedition of experiments on sodium with improved accuracy may help to clarify open theoretical many-body issues.

Based on the comparison with the Compton profile obtained by inelastic X-ray scattering spectroscopy experiments, non-self-consistent $G_0W_0$ turns out to be a very good approximation to calculate the momentum distribution and related quantities, while this is not yet evident for any attempt  including self-consistent GW, both on $G$ and $W$.
Therefore, we expect $G_0W_0$ to provide quantitative reference values for the momentum distribution valuable for the development of new exchange-correlation approximations within density-matrix functional theory.
Reverse  engineering from $G_0W_0$ to DMFT, is one of the promising ways to improve exchange-correlations in DMFT.
Similiar strategies have already successfully applied regarding improvement of time-dependent density-functional theory (TDDFT) by exploiting the Bethe-Salpeter equation.\cite{lrc,mt,Adragna,Marini,Tokatly,conskernel}.

Finally, it would be interesting to explore the limits of the GW approximation in reproducing the correct quasiparticle renormalization factor and the momentum distribution as measured in IXSS experiments on other systems.
An IXSS measure of such quantities, not influenced by Hartree or exchange effects like \eg the bandgap, would unambiguously assess the level of correlation in these systems and thus the validity of GW in describing correlations.
A still open question is whether GW can describe the quasiparticle renormalization factor in systems with increasing level of correlations, up to strongly correlated systems.

\section*{Acknowledgements}
Calculations were performed using our personal codes with checks using Mathematica for the jellium model, while for sodium we have used a modified version of \textsc{ABINIT}\cite{ABINIT} that will be made available in the next versions.
Fortran codes were run on IDRIS and Ciment/Phynum computers and at the supercomputer center of the Russian Academy of Sciences in Moscow.
Support was provided by the \textit{Fondation Nanosciences} via the NanoSTAR RTRA project.
A. T. is grateful to CNRS for a \textit{Charg\'e de Recherche Associ\'e} grant. 
S. H. and K. H. were supported by the Academy of Finland (contracts 1127462, 1256211 and 1254065) and the University of Helsinki research funds (contract 490076).

\bibliography{nagw}

\begin{thebibliography}{63}%
\makeatletter
\providecommand \@ifxundefined [1]{%
 \@ifx{#1\undefined}
}%
\providecommand \@ifnum [1]{%
 \ifnum #1\expandafter \@firstoftwo
 \else \expandafter \@secondoftwo
 \fi
}%
\providecommand \@ifx [1]{%
 \ifx #1\expandafter \@firstoftwo
 \else \expandafter \@secondoftwo
 \fi
}%
\providecommand \natexlab [1]{#1}%
\providecommand \enquote  [1]{``#1''}%
\providecommand \bibnamefont  [1]{#1}%
\providecommand \bibfnamefont [1]{#1}%
\providecommand \citenamefont [1]{#1}%
\providecommand \href@noop [0]{\@secondoftwo}%
\providecommand \href [0]{\begingroup \@sanitize@url \@href}%
\providecommand \@href[1]{\@@startlink{#1}\@@href}%
\providecommand \@@href[1]{\endgroup#1\@@endlink}%
\providecommand \@sanitize@url [0]{\catcode `\\12\catcode `\$12\catcode
  `\&12\catcode `\#12\catcode `\^12\catcode `\_12\catcode `\%12\relax}%
\providecommand \@@startlink[1]{}%
\providecommand \@@endlink[0]{}%
\providecommand \url  [0]{\begingroup\@sanitize@url \@url }%
\providecommand \@url [1]{\endgroup\@href {#1}{\urlprefix }}%
\providecommand \urlprefix  [0]{URL }%
\providecommand \Eprint [0]{\href }%
\providecommand \doibase [0]{http://dx.doi.org/}%
\providecommand \selectlanguage [0]{\@gobble}%
\providecommand \bibinfo  [0]{\@secondoftwo}%
\providecommand \bibfield  [0]{\@secondoftwo}%
\providecommand \translation [1]{[#1]}%
\providecommand \BibitemOpen [0]{}%
\providecommand \bibitemStop [0]{}%
\providecommand \bibitemNoStop [0]{.\EOS\space}%
\providecommand \EOS [0]{\spacefactor3000\relax}%
\providecommand \BibitemShut  [1]{\csname bibitem#1\endcsname}%
\let\auto@bib@innerbib\@empty
\bibitem [{\citenamefont {Nozi{\`e}res}(1997)}]{electrongas}%
  \BibitemOpen
  \bibfield  {author} {\bibinfo {author} {\bibfnamefont {P.}~\bibnamefont
  {Nozi{\`e}res}},\ }\href@noop {} {\emph {\bibinfo {title} {Theory of
  Interacting Fermi Systems}}}\ (\bibinfo  {publisher} {Westview Press},\
  \bibinfo {year} {1997})\BibitemShut {NoStop}%
\bibitem [{\citenamefont {Luttinger}(1960)}]{Luttinger60}%
  \BibitemOpen
  \bibfield  {author} {\bibinfo {author} {\bibfnamefont {J.~M.}\ \bibnamefont
  {Luttinger}},\ }\href@noop {} {\bibfield  {journal} {\bibinfo  {journal}
  {Phys. Rev.}\ }\textbf {\bibinfo {volume} {119}},\ \bibinfo {pages} {1153}
  (\bibinfo {year} {1960})}\BibitemShut {NoStop}%
\bibitem [{\citenamefont {Hedin}(1965{\natexlab{a}})}]{hedin65}%
  \BibitemOpen
  \bibfield  {author} {\bibinfo {author} {\bibfnamefont {L.}~\bibnamefont
  {Hedin}},\ }\href@noop {} {\bibfield  {journal} {\bibinfo  {journal} {Phys.
  Rev.}\ }\textbf {\bibinfo {volume} {139}},\ \bibinfo {pages} {A796} (\bibinfo
  {year} {1965}{\natexlab{a}})}\BibitemShut {NoStop}%
\bibitem [{\citenamefont {Rice}(1965)}]{rice65}%
  \BibitemOpen
  \bibfield  {author} {\bibinfo {author} {\bibfnamefont {T.}~\bibnamefont
  {Rice}},\ }\href@noop {} {\bibfield  {journal} {\bibinfo  {journal} {Ann.
  Phys. (N.Y.)}\ }\textbf {\bibinfo {volume} {31}},\ \bibinfo {pages} {100}
  (\bibinfo {year} {1965})}\BibitemShut {NoStop}%
\bibitem [{\citenamefont {Pajanne}\ and\ \citenamefont
  {Arponen}(1982)}]{pajanne82}%
  \BibitemOpen
  \bibfield  {author} {\bibinfo {author} {\bibfnamefont {E.}~\bibnamefont
  {Pajanne}}\ and\ \bibinfo {author} {\bibfnamefont {J.}~\bibnamefont
  {Arponen}},\ }\href@noop {} {\bibfield  {journal} {\bibinfo  {journal} {J.
  Phys. C: Solid State Phys.}\ }\textbf {\bibinfo {volume} {15}},\ \bibinfo
  {pages} {2683} (\bibinfo {year} {1982})}\BibitemShut {NoStop}%
\bibitem [{\citenamefont {Lam}(1971)}]{lam71}%
  \BibitemOpen
  \bibfield  {author} {\bibinfo {author} {\bibfnamefont {J.}~\bibnamefont
  {Lam}},\ }\href@noop {} {\bibfield  {journal} {\bibinfo  {journal} {Phys.
  Rev. B}\ }\textbf {\bibinfo {volume} {3}},\ \bibinfo {pages} {3243} (\bibinfo
  {year} {1971})}\BibitemShut {NoStop}%
\bibitem [{\citenamefont {Holm}\ and\ \citenamefont {{von
  Barth}}(1998)}]{holm98}%
  \BibitemOpen
  \bibfield  {author} {\bibinfo {author} {\bibfnamefont {B.}~\bibnamefont
  {Holm}}\ and\ \bibinfo {author} {\bibfnamefont {U.}~\bibnamefont {{von
  Barth}}},\ }\href@noop {} {\bibfield  {journal} {\bibinfo  {journal} {Phys.
  Rev. B}\ }\textbf {\bibinfo {volume} {57}},\ \bibinfo {pages} {2108}
  (\bibinfo {year} {1998})}\BibitemShut {NoStop}%
\bibitem [{\citenamefont {Lantto}(1980)}]{lantto80}%
  \BibitemOpen
  \bibfield  {author} {\bibinfo {author} {\bibfnamefont {L.~J.}\ \bibnamefont
  {Lantto}},\ }\href@noop {} {\bibfield  {journal} {\bibinfo  {journal} {Phys.
  Rev. B}\ }\textbf {\bibinfo {volume} {22}},\ \bibinfo {pages} {1380}
  (\bibinfo {year} {1980})}\BibitemShut {NoStop}%
\bibitem [{\citenamefont {Takada}\ and\ \citenamefont
  {Yasuhara}(1991)}]{takada91}%
  \BibitemOpen
  \bibfield  {author} {\bibinfo {author} {\bibfnamefont {Y.}~\bibnamefont
  {Takada}}\ and\ \bibinfo {author} {\bibfnamefont {H.}~\bibnamefont
  {Yasuhara}},\ }\href@noop {} {\bibfield  {journal} {\bibinfo  {journal}
  {Phys. Rev. B}\ }\textbf {\bibinfo {volume} {44}},\ \bibinfo {pages} {7879}
  (\bibinfo {year} {1991})}\BibitemShut {NoStop}%
\bibitem [{\citenamefont {Ziesche}(2002{\natexlab{a}})}]{Ziesche02b}%
  \BibitemOpen
  \bibfield  {author} {\bibinfo {author} {\bibfnamefont {P.}~\bibnamefont
  {Ziesche}},\ }\href@noop {} {\bibfield  {journal} {\bibinfo  {journal} {Phys.
  Status Solidi B}\ }\textbf {\bibinfo {volume} {232}},\ \bibinfo {pages} {231}
  (\bibinfo {year} {2002}{\natexlab{a}})}\BibitemShut {NoStop}%
\bibitem [{\citenamefont {Gori-Giorgi}\ and\ \citenamefont
  {Ziesche}(2002)}]{gori-giori02}%
  \BibitemOpen
  \bibfield  {author} {\bibinfo {author} {\bibfnamefont {P.}~\bibnamefont
  {Gori-Giorgi}}\ and\ \bibinfo {author} {\bibfnamefont {P.}~\bibnamefont
  {Ziesche}},\ }\href@noop {} {\bibfield  {journal} {\bibinfo  {journal} {Phys.
  Rev. B}\ }\textbf {\bibinfo {volume} {66}},\ \bibinfo {pages} {235116}
  (\bibinfo {year} {2002})}\BibitemShut {NoStop}%
\bibitem [{\citenamefont {Holzmann}\ \emph {et~al.}(2011)\citenamefont
  {Holzmann}, \citenamefont {Bernu}, \citenamefont {Pierleoni}, \citenamefont
  {McMinis}, \citenamefont {Ceperley}, \citenamefont {Olevano},\ and\
  \citenamefont {{Delle Site}}}]{Holzmannetal}%
  \BibitemOpen
  \bibfield  {author} {\bibinfo {author} {\bibfnamefont {M.}~\bibnamefont
  {Holzmann}}, \bibinfo {author} {\bibfnamefont {B.}~\bibnamefont {Bernu}},
  \bibinfo {author} {\bibfnamefont {C.}~\bibnamefont {Pierleoni}}, \bibinfo
  {author} {\bibfnamefont {J.}~\bibnamefont {McMinis}}, \bibinfo {author}
  {\bibfnamefont {D.~M.}\ \bibnamefont {Ceperley}}, \bibinfo {author}
  {\bibfnamefont {V.}~\bibnamefont {Olevano}}, \ and\ \bibinfo {author}
  {\bibfnamefont {L.}~\bibnamefont {{Delle Site}}},\ }\href@noop {} {\bibfield
  {journal} {\bibinfo  {journal} {Phys. Rev. Lett.}\ }\textbf {\bibinfo
  {volume} {107}},\ \bibinfo {pages} {110402} (\bibinfo {year}
  {2011})}\BibitemShut {NoStop}%
\bibitem [{\citenamefont {Maebashi}\ and\ \citenamefont
  {Takada}(2011)}]{Maebashi11}%
  \BibitemOpen
  \bibfield  {author} {\bibinfo {author} {\bibfnamefont {H.}~\bibnamefont
  {Maebashi}}\ and\ \bibinfo {author} {\bibfnamefont {Y.}~\bibnamefont
  {Takada}},\ }\href@noop {} {\bibfield  {journal} {\bibinfo  {journal} {Phys.
  Rev. B}\ }\textbf {\bibinfo {volume} {84}},\ \bibinfo {pages} {245134}
  (\bibinfo {year} {2011})}\BibitemShut {NoStop}%
\bibitem [{\citenamefont {Huotari}\ \emph {et~al.}(2010)\citenamefont {Huotari}
  \emph {et~al.}}]{Huotarietal}%
  \BibitemOpen
  \bibfield  {author} {\bibinfo {author} {\bibfnamefont {S.}~\bibnamefont
  {Huotari}} \emph {et~al.},\ }\href@noop {} {\bibfield  {journal} {\bibinfo
  {journal} {Phys. Rev. Lett.}\ }\textbf {\bibinfo {volume} {105}},\ \bibinfo
  {pages} {086403} (\bibinfo {year} {2010})}\BibitemShut {NoStop}%
\bibitem [{\citenamefont {Lodwin}(1955)}]{Lodwin55}%
  \BibitemOpen
  \bibfield  {author} {\bibinfo {author} {\bibfnamefont {P.~O.}\ \bibnamefont
  {Lodwin}},\ }\href@noop {} {\bibfield  {journal} {\bibinfo  {journal} {Phys.
  Rev.}\ }\textbf {\bibinfo {volume} {97}},\ \bibinfo {pages} {1474} (\bibinfo
  {year} {1955})}\BibitemShut {NoStop}%
\bibitem [{\citenamefont {Gilbert}(1975)}]{Gilbert75}%
  \BibitemOpen
  \bibfield  {author} {\bibinfo {author} {\bibfnamefont {T.~L.}\ \bibnamefont
  {Gilbert}},\ }\href@noop {} {\bibfield  {journal} {\bibinfo  {journal} {Phys.
  Rev. B}\ }\textbf {\bibinfo {volume} {12}},\ \bibinfo {pages} {2111}
  (\bibinfo {year} {1975})}\BibitemShut {NoStop}%
\bibitem [{\citenamefont {Ziesche}(2002{\natexlab{b}})}]{Ziesche02}%
  \BibitemOpen
  \bibfield  {author} {\bibinfo {author} {\bibfnamefont {P.}~\bibnamefont
  {Ziesche}},\ }\href@noop {} {\bibfield  {journal} {\bibinfo  {journal} {Int.
  J. Quantum Chem.}\ }\textbf {\bibinfo {volume} {90}},\ \bibinfo {pages} {342}
  (\bibinfo {year} {2002}{\natexlab{b}})}\BibitemShut {NoStop}%
\bibitem [{\citenamefont {Kubo}(1996)}]{Kubo96}%
  \BibitemOpen
  \bibfield  {author} {\bibinfo {author} {\bibfnamefont {Y.}~\bibnamefont
  {Kubo}},\ }\href@noop {} {\bibfield  {journal} {\bibinfo  {journal} {J. Phys.
  Soc. Jpn.}\ }\textbf {\bibinfo {volume} {65}},\ \bibinfo {pages} {16}
  (\bibinfo {year} {1996})}\BibitemShut {NoStop}%
\bibitem [{\citenamefont {Kubo}(1997)}]{Kubo97}%
  \BibitemOpen
  \bibfield  {author} {\bibinfo {author} {\bibfnamefont {Y.}~\bibnamefont
  {Kubo}},\ }\href@noop {} {\bibfield  {journal} {\bibinfo  {journal} {J. Phys.
  Soc. Jpn.}\ }\textbf {\bibinfo {volume} {66}},\ \bibinfo {pages} {2236}
  (\bibinfo {year} {1997})}\BibitemShut {NoStop}%
\bibitem [{\citenamefont {Kubo}(2001)}]{Kubo01}%
  \BibitemOpen
  \bibfield  {author} {\bibinfo {author} {\bibfnamefont {Y.}~\bibnamefont
  {Kubo}},\ }\href@noop {} {\bibfield  {journal} {\bibinfo  {journal} {J. Phys.
  Chem. Solids}\ }\textbf {\bibinfo {volume} {62}},\ \bibinfo {pages} {2199}
  (\bibinfo {year} {2001})}\BibitemShut {NoStop}%
\bibitem [{\citenamefont {Hamada}\ \emph {et~al.}(1990)\citenamefont {Hamada},
  \citenamefont {Hwang},\ and\ \citenamefont {Freeman}}]{HamadaHwangFreeman90}%
  \BibitemOpen
  \bibfield  {author} {\bibinfo {author} {\bibfnamefont {N.}~\bibnamefont
  {Hamada}}, \bibinfo {author} {\bibfnamefont {M.}~\bibnamefont {Hwang}}, \
  and\ \bibinfo {author} {\bibfnamefont {A.~J.}\ \bibnamefont {Freeman}},\
  }\href@noop {} {\bibfield  {journal} {\bibinfo  {journal} {Phys. Rev. B}\
  }\textbf {\bibinfo {volume} {41}},\ \bibinfo {pages} {3620} (\bibinfo {year}
  {1990})}\BibitemShut {NoStop}%
\bibitem [{\citenamefont {Sch{\"u}lke}(1997)}]{Schuelke97}%
  \BibitemOpen
  \bibfield  {author} {\bibinfo {author} {\bibfnamefont {W.}~\bibnamefont
  {Sch{\"u}lke}},\ }\href@noop {} {\bibfield  {journal} {\bibinfo  {journal}
  {J. Phys. Soc. Jpn.}\ }\textbf {\bibinfo {volume} {66}},\ \bibinfo {pages}
  {2470} (\bibinfo {year} {1997})}\BibitemShut {NoStop}%
\bibitem [{\citenamefont {Huotari}\ \emph {et~al.}(2002)\citenamefont
  {Huotari}, \citenamefont {H{\"a}m{\"a}l{\"a}inen}, \citenamefont {Manninen},
  \citenamefont {Sternemann}, \citenamefont {Kaprolat}, \citenamefont
  {Sch{\"u}lke},\ and\ \citenamefont {Buslaps}}]{huotari2002}%
  \BibitemOpen
  \bibfield  {author} {\bibinfo {author} {\bibfnamefont {S.}~\bibnamefont
  {Huotari}}, \bibinfo {author} {\bibfnamefont {K.}~\bibnamefont
  {H{\"a}m{\"a}l{\"a}inen}}, \bibinfo {author} {\bibfnamefont {S.}~\bibnamefont
  {Manninen}}, \bibinfo {author} {\bibfnamefont {C.}~\bibnamefont
  {Sternemann}}, \bibinfo {author} {\bibfnamefont {A.}~\bibnamefont
  {Kaprolat}}, \bibinfo {author} {\bibfnamefont {W.}~\bibnamefont
  {Sch{\"u}lke}}, \ and\ \bibinfo {author} {\bibfnamefont {T.}~\bibnamefont
  {Buslaps}},\ }\href@noop {} {\bibfield  {journal} {\bibinfo  {journal} {Phys.
  Rev. B}\ }\textbf {\bibinfo {volume} {66}},\ \bibinfo {pages} {085104}
  (\bibinfo {year} {2002})}\BibitemShut {NoStop}%
\bibitem [{\citenamefont {Sakurai}\ \emph {et~al.}(2011)\citenamefont
  {Sakurai}, \citenamefont {Itou}, \citenamefont {Barbiellini}, \citenamefont
  {Mijnarends}, \citenamefont {Markiewicz}, \citenamefont {Kaprzyk},
  \citenamefont {Gillet}, \citenamefont {Wakimoto}, \citenamefont {Fujita},
  \citenamefont {Basak}, \citenamefont {Wang}, \citenamefont {Al-Sawai},
  \citenamefont {Lin}, \citenamefont {Bansil},\ and\ \citenamefont
  {Yamada}}]{sakurai2011}%
  \BibitemOpen
  \bibfield  {author} {\bibinfo {author} {\bibfnamefont {Y.}~\bibnamefont
  {Sakurai}}, \bibinfo {author} {\bibfnamefont {M.}~\bibnamefont {Itou}},
  \bibinfo {author} {\bibfnamefont {B.}~\bibnamefont {Barbiellini}}, \bibinfo
  {author} {\bibfnamefont {P.~E.}\ \bibnamefont {Mijnarends}}, \bibinfo
  {author} {\bibfnamefont {R.~S.}\ \bibnamefont {Markiewicz}}, \bibinfo
  {author} {\bibfnamefont {S.}~\bibnamefont {Kaprzyk}}, \bibinfo {author}
  {\bibfnamefont {J.-M.}\ \bibnamefont {Gillet}}, \bibinfo {author}
  {\bibfnamefont {S.}~\bibnamefont {Wakimoto}}, \bibinfo {author}
  {\bibfnamefont {M.}~\bibnamefont {Fujita}}, \bibinfo {author} {\bibfnamefont
  {S.}~\bibnamefont {Basak}}, \bibinfo {author} {\bibfnamefont {Y.-J.}\
  \bibnamefont {Wang}}, \bibinfo {author} {\bibfnamefont {W.}~\bibnamefont
  {Al-Sawai}}, \bibinfo {author} {\bibfnamefont {H.}~\bibnamefont {Lin}},
  \bibinfo {author} {\bibfnamefont {A.}~\bibnamefont {Bansil}}, \ and\ \bibinfo
  {author} {\bibfnamefont {K.}~\bibnamefont {Yamada}},\ }\href@noop {}
  {\bibfield  {journal} {\bibinfo  {journal} {Science}\ }\textbf {\bibinfo
  {volume} {332}},\ \bibinfo {pages} {698} (\bibinfo {year}
  {2011})}\BibitemShut {NoStop}%
\bibitem [{\citenamefont {Huotari}\ \emph {et~al.}(2009)\citenamefont
  {Huotari}, \citenamefont {Boldrini}, \citenamefont {Honkim{\"a}ki},
  \citenamefont {Suortti},\ and\ \citenamefont {Weyrich}}]{huotari2009}%
  \BibitemOpen
  \bibfield  {author} {\bibinfo {author} {\bibfnamefont {S.}~\bibnamefont
  {Huotari}}, \bibinfo {author} {\bibfnamefont {B.}~\bibnamefont {Boldrini}},
  \bibinfo {author} {\bibfnamefont {V.}~\bibnamefont {Honkim{\"a}ki}}, \bibinfo
  {author} {\bibfnamefont {P.}~\bibnamefont {Suortti}}, \ and\ \bibinfo
  {author} {\bibfnamefont {W.}~\bibnamefont {Weyrich}},\ }\href@noop {}
  {\bibfield  {journal} {\bibinfo  {journal} {J. Synchrotron Radiat.}\ }\textbf
  {\bibinfo {volume} {16}},\ \bibinfo {pages} {672} (\bibinfo {year}
  {2009})}\BibitemShut {NoStop}%
\bibitem [{\citenamefont {Kontrym-Sznajd}\ \emph {et~al.}(2003)\citenamefont
  {Kontrym-Sznajd}, \citenamefont {Samsel-Czekala}, \citenamefont {Huotari},
  \citenamefont {H{\"a}m{\"a}l{\"a}inen},\ and\ \citenamefont
  {Manninen}}]{kontrym2003}%
  \BibitemOpen
  \bibfield  {author} {\bibinfo {author} {\bibfnamefont {G.}~\bibnamefont
  {Kontrym-Sznajd}}, \bibinfo {author} {\bibfnamefont {M.}~\bibnamefont
  {Samsel-Czekala}}, \bibinfo {author} {\bibfnamefont {S.}~\bibnamefont
  {Huotari}}, \bibinfo {author} {\bibfnamefont {K.}~\bibnamefont
  {H{\"a}m{\"a}l{\"a}inen}}, \ and\ \bibinfo {author} {\bibfnamefont
  {S.}~\bibnamefont {Manninen}},\ }\href@noop {} {\bibfield  {journal}
  {\bibinfo  {journal} {Phys. Rev. B}\ }\textbf {\bibinfo {volume} {68}},\
  \bibinfo {pages} {155106} (\bibinfo {year} {2003})}\BibitemShut {NoStop}%
\bibitem [{\citenamefont {Kontrym-Sznajd}\ \emph {et~al.}(2002)\citenamefont
  {Kontrym-Sznajd}, \citenamefont {Samsel-Czekala}, \citenamefont {Pietraszko},
  \citenamefont {Sormann}, \citenamefont {Manninen}, \citenamefont {Huotari},
  \citenamefont {H{\"a}m{\"a}l{\"a}inen}, \citenamefont {Laukkanen},
  \citenamefont {West},\ and\ \citenamefont {Sch{\"u}lke}}]{kontrym2002}%
  \BibitemOpen
  \bibfield  {author} {\bibinfo {author} {\bibfnamefont {G.}~\bibnamefont
  {Kontrym-Sznajd}}, \bibinfo {author} {\bibfnamefont {M.}~\bibnamefont
  {Samsel-Czekala}}, \bibinfo {author} {\bibfnamefont {A.}~\bibnamefont
  {Pietraszko}}, \bibinfo {author} {\bibfnamefont {H.}~\bibnamefont {Sormann}},
  \bibinfo {author} {\bibfnamefont {S.}~\bibnamefont {Manninen}}, \bibinfo
  {author} {\bibfnamefont {S.}~\bibnamefont {Huotari}}, \bibinfo {author}
  {\bibfnamefont {K.}~\bibnamefont {H{\"a}m{\"a}l{\"a}inen}}, \bibinfo {author}
  {\bibfnamefont {J.}~\bibnamefont {Laukkanen}}, \bibinfo {author}
  {\bibfnamefont {R.~N.}\ \bibnamefont {West}}, \ and\ \bibinfo {author}
  {\bibfnamefont {W.}~\bibnamefont {Sch{\"u}lke}},\ }\href@noop {} {\bibfield
  {journal} {\bibinfo  {journal} {Phys. Rev. B}\ }\textbf {\bibinfo {volume}
  {66}},\ \bibinfo {pages} {155110} (\bibinfo {year} {2002})}\BibitemShut
  {NoStop}%
\bibitem [{\citenamefont {Huotari}\ \emph {et~al.}(2000)\citenamefont
  {Huotari}, \citenamefont {H{\"a}m{\"a}l{\"a}inen}, \citenamefont {Manninen},
  \citenamefont {Kaprzyk}, \citenamefont {Bansil}, \citenamefont {Caliebe},
  \citenamefont {Buslaps}, \citenamefont {Honkim{\"a}ki},\ and\ \citenamefont
  {Suortti}}]{huotari2000}%
  \BibitemOpen
  \bibfield  {author} {\bibinfo {author} {\bibfnamefont {S.}~\bibnamefont
  {Huotari}}, \bibinfo {author} {\bibfnamefont {K.}~\bibnamefont
  {H{\"a}m{\"a}l{\"a}inen}}, \bibinfo {author} {\bibfnamefont {S.}~\bibnamefont
  {Manninen}}, \bibinfo {author} {\bibfnamefont {S.}~\bibnamefont {Kaprzyk}},
  \bibinfo {author} {\bibfnamefont {A.}~\bibnamefont {Bansil}}, \bibinfo
  {author} {\bibfnamefont {W.}~\bibnamefont {Caliebe}}, \bibinfo {author}
  {\bibfnamefont {T.}~\bibnamefont {Buslaps}}, \bibinfo {author} {\bibfnamefont
  {V.}~\bibnamefont {Honkim{\"a}ki}}, \ and\ \bibinfo {author} {\bibfnamefont
  {P.}~\bibnamefont {Suortti}},\ }\href@noop {} {\bibfield  {journal} {\bibinfo
   {journal} {Phys. Rev. B}\ }\textbf {\bibinfo {volume} {62}},\ \bibinfo
  {pages} {7956} (\bibinfo {year} {2000})}\BibitemShut {NoStop}%
\bibitem [{\citenamefont {Holzmann}\ \emph {et~al.}(2009)\citenamefont
  {Holzmann}, \citenamefont {Bernu}, \citenamefont {Olevano}, \citenamefont
  {Martin},\ and\ \citenamefont {Ceperley}}]{holzmann09}%
  \BibitemOpen
  \bibfield  {author} {\bibinfo {author} {\bibfnamefont {M.}~\bibnamefont
  {Holzmann}}, \bibinfo {author} {\bibfnamefont {B.}~\bibnamefont {Bernu}},
  \bibinfo {author} {\bibfnamefont {V.}~\bibnamefont {Olevano}}, \bibinfo
  {author} {\bibfnamefont {R.~M.}\ \bibnamefont {Martin}}, \ and\ \bibinfo
  {author} {\bibfnamefont {D.~M.}\ \bibnamefont {Ceperley}},\ }\href@noop {}
  {\bibfield  {journal} {\bibinfo  {journal} {Phys. Rev. B}\ }\textbf {\bibinfo
  {volume} {79}},\ \bibinfo {pages} {041308(R)} (\bibinfo {year}
  {2009})}\BibitemShut {NoStop}%
\bibitem [{\citenamefont {Cooper}\ \emph {et~al.}(2004)\citenamefont {Cooper},
  \citenamefont {Mijnarends}, \citenamefont {Shiotani}, \citenamefont {Sakai},\
  and\ \citenamefont {Bansil}}]{cooperbook}%
  \BibitemOpen
  \bibinfo {editor} {\bibfnamefont {M.~J.}\ \bibnamefont {Cooper}}, \bibinfo
  {editor} {\bibfnamefont {P.~E.}\ \bibnamefont {Mijnarends}}, \bibinfo
  {editor} {\bibfnamefont {N.}~\bibnamefont {Shiotani}}, \bibinfo {editor}
  {\bibfnamefont {N.}~\bibnamefont {Sakai}}, \ and\ \bibinfo {editor}
  {\bibfnamefont {A.}~\bibnamefont {Bansil}},\ eds.,\ \href@noop {} {\emph
  {\bibinfo {title} {X-Ray Compton Scattering}}}\ (\bibinfo  {publisher}
  {Oxford University Press, Oxford},\ \bibinfo {year} {2004})\BibitemShut
  {NoStop}%
\bibitem [{\citenamefont {Eisenberger}\ and\ \citenamefont
  {Platzman}(1970)}]{eisenberger1970}%
  \BibitemOpen
  \bibfield  {author} {\bibinfo {author} {\bibfnamefont {P.}~\bibnamefont
  {Eisenberger}}\ and\ \bibinfo {author} {\bibfnamefont {P.~M.}\ \bibnamefont
  {Platzman}},\ }\href@noop {} {\bibfield  {journal} {\bibinfo  {journal}
  {Phys. Rev. A}\ }\textbf {\bibinfo {volume} {2}},\ \bibinfo {pages} {415}
  (\bibinfo {year} {1970})}\BibitemShut {NoStop}%
\bibitem [{\citenamefont {Verbeni}\ \emph {et~al.}(2009)\citenamefont
  {Verbeni}, \citenamefont {Pylkk{\"a}nen}, \citenamefont {Huotari},
  \citenamefont {Simonelli}, \citenamefont {Vank{\'o}}, \citenamefont {Martel},
  \citenamefont {Henriquet},\ and\ \citenamefont {Monaco}}]{verbeni2009}%
  \BibitemOpen
  \bibfield  {author} {\bibinfo {author} {\bibfnamefont {R.}~\bibnamefont
  {Verbeni}}, \bibinfo {author} {\bibfnamefont {T.}~\bibnamefont
  {Pylkk{\"a}nen}}, \bibinfo {author} {\bibfnamefont {S.}~\bibnamefont
  {Huotari}}, \bibinfo {author} {\bibfnamefont {L.}~\bibnamefont {Simonelli}},
  \bibinfo {author} {\bibfnamefont {G.}~\bibnamefont {Vank{\'o}}}, \bibinfo
  {author} {\bibfnamefont {K.}~\bibnamefont {Martel}}, \bibinfo {author}
  {\bibfnamefont {C.}~\bibnamefont {Henriquet}}, \ and\ \bibinfo {author}
  {\bibfnamefont {G.}~\bibnamefont {Monaco}},\ }\href@noop {} {\bibfield
  {journal} {\bibinfo  {journal} {J. Synchrotron Radiat.}\ }\textbf {\bibinfo
  {volume} {16}},\ \bibinfo {pages} {469} (\bibinfo {year} {2009})}\BibitemShut
  {NoStop}%
\bibitem [{\citenamefont {Huotari}\ \emph {et~al.}(2005)\citenamefont
  {Huotari}, \citenamefont {Vank{\'o}}, \citenamefont {Albergamo},
  \citenamefont {Ponchut}, \citenamefont {Graafsma}, \citenamefont {Henriquet},
  \citenamefont {Verbeni},\ and\ \citenamefont {Monaco}}]{huotari2005}%
  \BibitemOpen
  \bibfield  {author} {\bibinfo {author} {\bibfnamefont {S.}~\bibnamefont
  {Huotari}}, \bibinfo {author} {\bibfnamefont {G.}~\bibnamefont {Vank{\'o}}},
  \bibinfo {author} {\bibfnamefont {F.}~\bibnamefont {Albergamo}}, \bibinfo
  {author} {\bibfnamefont {C.}~\bibnamefont {Ponchut}}, \bibinfo {author}
  {\bibfnamefont {H.}~\bibnamefont {Graafsma}}, \bibinfo {author}
  {\bibfnamefont {C.}~\bibnamefont {Henriquet}}, \bibinfo {author}
  {\bibfnamefont {R.}~\bibnamefont {Verbeni}}, \ and\ \bibinfo {author}
  {\bibfnamefont {G.}~\bibnamefont {Monaco}},\ }\href@noop {} {\bibfield
  {journal} {\bibinfo  {journal} {J. Synchrotron Radiat.}\ }\textbf {\bibinfo
  {volume} {12}},\ \bibinfo {pages} {467} (\bibinfo {year} {2005})}\BibitemShut
  {NoStop}%
\bibitem [{\citenamefont {Issolah}\ \emph {et~al.}(1988)\citenamefont
  {Issolah}, \citenamefont {L{\'e}vy}, \citenamefont {Beswick},\ and\
  \citenamefont {Loupias}}]{issolah1988}%
  \BibitemOpen
  \bibfield  {author} {\bibinfo {author} {\bibfnamefont {A.}~\bibnamefont
  {Issolah}}, \bibinfo {author} {\bibfnamefont {B.}~\bibnamefont {L{\'e}vy}},
  \bibinfo {author} {\bibfnamefont {A.}~\bibnamefont {Beswick}}, \ and\
  \bibinfo {author} {\bibfnamefont {G.}~\bibnamefont {Loupias}},\ }\href@noop
  {} {\bibfield  {journal} {\bibinfo  {journal} {Phys. Rev. A}\ }\textbf
  {\bibinfo {volume} {38}},\ \bibinfo {pages} {4509} (\bibinfo {year}
  {1988})}\BibitemShut {NoStop}%
\bibitem [{\citenamefont {Soininen}\ \emph {et~al.}(2005)\citenamefont
  {Soininen}, \citenamefont {Ankudinov},\ and\ \citenamefont
  {Rehr}}]{soininen2005}%
  \BibitemOpen
  \bibfield  {author} {\bibinfo {author} {\bibfnamefont {J.~A.}\ \bibnamefont
  {Soininen}}, \bibinfo {author} {\bibfnamefont {A.~L.}\ \bibnamefont
  {Ankudinov}}, \ and\ \bibinfo {author} {\bibfnamefont {J.~J.}\ \bibnamefont
  {Rehr}},\ }\href@noop {} {\bibfield  {journal} {\bibinfo  {journal} {Phys.
  Rev. B}\ }\textbf {\bibinfo {volume} {72}},\ \bibinfo {pages} {045136}
  (\bibinfo {year} {2005})}\BibitemShut {NoStop}%
\bibitem [{\citenamefont {Hybertsen}\ and\ \citenamefont
  {Louie}(1985)}]{HybertsenLouie}%
  \BibitemOpen
  \bibfield  {author} {\bibinfo {author} {\bibfnamefont {M.~S.}\ \bibnamefont
  {Hybertsen}}\ and\ \bibinfo {author} {\bibfnamefont {S.~G.}\ \bibnamefont
  {Louie}},\ }\href@noop {} {\bibfield  {journal} {\bibinfo  {journal} {Phys.
  Rev. Lett.}\ }\textbf {\bibinfo {volume} {55}},\ \bibinfo {pages} {1418}
  (\bibinfo {year} {1985})}\BibitemShut {NoStop}%
\bibitem [{\citenamefont {Gatti}\ \emph {et~al.}(2007)\citenamefont {Gatti},
  \citenamefont {Bruneval}, \citenamefont {Olevano},\ and\ \citenamefont
  {Reining}}]{GattiBrunevalOlevano}%
  \BibitemOpen
  \bibfield  {author} {\bibinfo {author} {\bibfnamefont {M.}~\bibnamefont
  {Gatti}}, \bibinfo {author} {\bibfnamefont {F.}~\bibnamefont {Bruneval}},
  \bibinfo {author} {\bibfnamefont {V.}~\bibnamefont {Olevano}}, \ and\
  \bibinfo {author} {\bibfnamefont {L.}~\bibnamefont {Reining}},\ }\href@noop
  {} {\bibfield  {journal} {\bibinfo  {journal} {Phys. Rev. Lett.}\ }\textbf
  {\bibinfo {volume} {99}},\ \bibinfo {pages} {266402} (\bibinfo {year}
  {2007})}\BibitemShut {NoStop}%
\bibitem [{\citenamefont {Botti}\ \emph {et~al.}(2002)\citenamefont {Botti}
  \emph {et~al.}}]{anisotropy}%
  \BibitemOpen
  \bibfield  {author} {\bibinfo {author} {\bibfnamefont {S.}~\bibnamefont
  {Botti}} \emph {et~al.},\ }\href@noop {} {\bibfield  {journal} {\bibinfo
  {journal} {Phys. Rev. Lett.}\ }\textbf {\bibinfo {volume} {89}},\ \bibinfo
  {pages} {216803} (\bibinfo {year} {2002})}\BibitemShut {NoStop}%
\bibitem [{\citenamefont {Hedin}(1965{\natexlab{b}})}]{Hedin}%
  \BibitemOpen
  \bibfield  {author} {\bibinfo {author} {\bibfnamefont {L.}~\bibnamefont
  {Hedin}},\ }\href@noop {} {\bibfield  {journal} {\bibinfo  {journal} {Phys.
  Rev.}\ }\textbf {\bibinfo {volume} {139}},\ \bibinfo {pages} {A796} (\bibinfo
  {year} {1965}{\natexlab{b}})}\BibitemShut {NoStop}%
\bibitem [{\citenamefont {Lindhard}(1954)}]{Lindhard}%
  \BibitemOpen
  \bibfield  {author} {\bibinfo {author} {\bibfnamefont {J.}~\bibnamefont
  {Lindhard}},\ }\href@noop {} {\bibfield  {journal} {\bibinfo  {journal} {Kgl.
  Danske Videnskab. Selskab, Mat. Fys. Medd.}\ }\textbf {\bibinfo {volume}
  {28}},\ \bibinfo {pages} {No. 8} (\bibinfo {year} {1954})}\BibitemShut
  {NoStop}%
\bibitem [{\citenamefont {Aulbur}\ \emph {et~al.}(1999)\citenamefont {Aulbur},
  \citenamefont {J{\"o}nsson},\ and\ \citenamefont {Wilkins}}]{AulburJonsson}%
  \BibitemOpen
  \bibfield  {author} {\bibinfo {author} {\bibfnamefont {W.}~\bibnamefont
  {Aulbur}}, \bibinfo {author} {\bibfnamefont {L.}~\bibnamefont {J{\"o}nsson}},
  \ and\ \bibinfo {author} {\bibfnamefont {J.~W.}\ \bibnamefont {Wilkins}},\
  }\href@noop {} {\bibfield  {journal} {\bibinfo  {journal} {Solid State
  Physics}\ }\textbf {\bibinfo {volume} {54}},\ \bibinfo {pages} {1} (\bibinfo
  {year} {1999})}\BibitemShut {NoStop}%
\bibitem [{\citenamefont {van Schilfgaarde}\ \emph {et~al.}(2006)\citenamefont
  {van Schilfgaarde}, \citenamefont {Kotani},\ and\ \citenamefont
  {Faleev}}]{Kotani}%
  \BibitemOpen
  \bibfield  {author} {\bibinfo {author} {\bibfnamefont {M.}~\bibnamefont {van
  Schilfgaarde}}, \bibinfo {author} {\bibfnamefont {T.}~\bibnamefont {Kotani}},
  \ and\ \bibinfo {author} {\bibfnamefont {S.}~\bibnamefont {Faleev}},\
  }\href@noop {} {\bibfield  {journal} {\bibinfo  {journal} {Phys. Rev. Lett.}\
  }\textbf {\bibinfo {volume} {96}},\ \bibinfo {pages} {226402} (\bibinfo
  {year} {2006})}\BibitemShut {NoStop}%
\bibitem [{\citenamefont {Bruneval}\ \emph {et~al.}(2006)\citenamefont
  {Bruneval}, \citenamefont {Vast},\ and\ \citenamefont {Reining}}]{Bruneval}%
  \BibitemOpen
  \bibfield  {author} {\bibinfo {author} {\bibfnamefont {F.}~\bibnamefont
  {Bruneval}}, \bibinfo {author} {\bibfnamefont {N.}~\bibnamefont {Vast}}, \
  and\ \bibinfo {author} {\bibfnamefont {L.}~\bibnamefont {Reining}},\
  }\href@noop {} {\bibfield  {journal} {\bibinfo  {journal} {Phys. Rev. B}\
  }\textbf {\bibinfo {volume} {74}},\ \bibinfo {pages} {045102} (\bibinfo
  {year} {2006})}\BibitemShut {NoStop}%
\bibitem [{\citenamefont {Holm}\ and\ \citenamefont {von
  Barth}(1998)}]{HolmVonBarth}%
  \BibitemOpen
  \bibfield  {author} {\bibinfo {author} {\bibfnamefont {B.}~\bibnamefont
  {Holm}}\ and\ \bibinfo {author} {\bibfnamefont {U.}~\bibnamefont {von
  Barth}},\ }\href@noop {} {\bibfield  {journal} {\bibinfo  {journal} {Phys.
  Rev. B}\ }\textbf {\bibinfo {volume} {57}},\ \bibinfo {pages} {2108}
  (\bibinfo {year} {1998})}\BibitemShut {NoStop}%
\bibitem [{\citenamefont {von Barth}\ and\ \citenamefont
  {Holm}(1996)}]{VonBarthHolm}%
  \BibitemOpen
  \bibfield  {author} {\bibinfo {author} {\bibfnamefont {U.}~\bibnamefont {von
  Barth}}\ and\ \bibinfo {author} {\bibfnamefont {B.}~\bibnamefont {Holm}},\
  }\href@noop {} {\bibfield  {journal} {\bibinfo  {journal} {Phys. Rev. B}\
  }\textbf {\bibinfo {volume} {54}},\ \bibinfo {pages} {8411} (\bibinfo {year}
  {1996})}\BibitemShut {NoStop}%
\bibitem [{\citenamefont {Lundqvist}(1967)}]{Lundqvist1967}%
  \BibitemOpen
  \bibfield  {author} {\bibinfo {author} {\bibfnamefont {B.~I.}\ \bibnamefont
  {Lundqvist}},\ }\href@noop {} {\bibfield  {journal} {\bibinfo  {journal}
  {Phys. Kondens. Materie}\ }\textbf {\bibinfo {volume} {6}},\ \bibinfo {pages}
  {206} (\bibinfo {year} {1967})}\BibitemShut {NoStop}%
\bibitem [{\citenamefont {Lundqvist}(1968)}]{Lundqvist1968}%
  \BibitemOpen
  \bibfield  {author} {\bibinfo {author} {\bibfnamefont {B.~I.}\ \bibnamefont
  {Lundqvist}},\ }\href@noop {} {\bibfield  {journal} {\bibinfo  {journal}
  {Phys. Kondens. Materie}\ }\textbf {\bibinfo {volume} {7}},\ \bibinfo {pages}
  {117} (\bibinfo {year} {1968})}\BibitemShut {NoStop}%
\bibitem [{Note1()}]{Note1}%
  \BibitemOpen
  \bibinfo {note} {It should be noticed that Eq.~(\ref {np1-A}) is equivalent
  to Eq.~(\ref {npA}) only in the exact theory or for number-conserving
  approximations. Non-self-consistent $G_0W_0$ is not conserving. However, the
  difference one can find in $G_0W_0$ between the two formulas cannot be larger
  than 0.001 at metallic densities, as it has been found by Schindlmayr
  \protect \textit {et al.}\cite {Schindlmayr01} This value is not appreciable
  here due to our numerical error which is not smaller than 0.01 both in the
  real sodium and in the jellium calculations.}\BibitemShut {Stop}%
\bibitem [{\citenamefont {Lundqvist}\ and\ \citenamefont
  {Lyd{\'e}n}(1971)}]{LundqvistLyden71}%
  \BibitemOpen
  \bibfield  {author} {\bibinfo {author} {\bibfnamefont {B.~I.}\ \bibnamefont
  {Lundqvist}}\ and\ \bibinfo {author} {\bibfnamefont {C.}~\bibnamefont
  {Lyd{\'e}n}},\ }\href@noop {} {\bibfield  {journal} {\bibinfo  {journal}
  {Phys. Rev. B}\ }\textbf {\bibinfo {volume} {4}},\ \bibinfo {pages} {3360}
  (\bibinfo {year} {1971})}\BibitemShut {NoStop}%
\bibitem [{\citenamefont {{Del Sole}}\ \emph {et~al.}(2003)\citenamefont {{Del
  Sole}}, \citenamefont {Adragna}, \citenamefont {Olevano},\ and\ \citenamefont
  {Reining}}]{dynamicaleffects}%
  \BibitemOpen
  \bibfield  {author} {\bibinfo {author} {\bibfnamefont {R.}~\bibnamefont {{Del
  Sole}}}, \bibinfo {author} {\bibfnamefont {G.}~\bibnamefont {Adragna}},
  \bibinfo {author} {\bibfnamefont {V.}~\bibnamefont {Olevano}}, \ and\
  \bibinfo {author} {\bibfnamefont {L.}~\bibnamefont {Reining}},\ }\href@noop
  {} {\bibfield  {journal} {\bibinfo  {journal} {Phys. Rev. B}\ }\textbf
  {\bibinfo {volume} {67}},\ \bibinfo {pages} {045207} (\bibinfo {year}
  {2003})}\BibitemShut {NoStop}%
\bibitem [{\citenamefont {Filippi}\ and\ \citenamefont
  {Ceperley}(1999)}]{Filippi99}%
  \BibitemOpen
  \bibfield  {author} {\bibinfo {author} {\bibfnamefont {C.}~\bibnamefont
  {Filippi}}\ and\ \bibinfo {author} {\bibfnamefont {D.~M.}\ \bibnamefont
  {Ceperley}},\ }\href@noop {} {\bibfield  {journal} {\bibinfo  {journal}
  {Phys. Rev. B}\ }\textbf {\bibinfo {volume} {59}},\ \bibinfo {pages} {7907}
  (\bibinfo {year} {1999})}\BibitemShut {NoStop}%
\bibitem [{\citenamefont {Tanaka}\ \emph {et~al.}(2001)\citenamefont {Tanaka}
  \emph {et~al.}}]{Liexp1}%
  \BibitemOpen
  \bibfield  {author} {\bibinfo {author} {\bibfnamefont {Y.}~\bibnamefont
  {Tanaka}} \emph {et~al.},\ }\href@noop {} {\bibfield  {journal} {\bibinfo
  {journal} {Phys. Rev. B}\ }\textbf {\bibinfo {volume} {63}},\ \bibinfo
  {pages} {045120} (\bibinfo {year} {2001})}\BibitemShut {NoStop}%
\bibitem [{\citenamefont {Sch{\" u}lke}\ \emph {et~al.}(1996)\citenamefont
  {Sch{\" u}lke}, \citenamefont {Stutz}, \citenamefont {Wohlert},\ and\
  \citenamefont {Kaprolat}}]{liexp2}%
  \BibitemOpen
  \bibfield  {author} {\bibinfo {author} {\bibfnamefont {W.}~\bibnamefont
  {Sch{\" u}lke}}, \bibinfo {author} {\bibfnamefont {G.}~\bibnamefont {Stutz}},
  \bibinfo {author} {\bibfnamefont {F.}~\bibnamefont {Wohlert}}, \ and\
  \bibinfo {author} {\bibfnamefont {A.}~\bibnamefont {Kaprolat}},\ }\href@noop
  {} {\bibfield  {journal} {\bibinfo  {journal} {Phys. Rev. B}\ }\textbf
  {\bibinfo {volume} {54}},\ \bibinfo {pages} {14381} (\bibinfo {year}
  {1996})}\BibitemShut {NoStop}%
\bibitem [{\citenamefont {Huotari}\ \emph {et~al.}(2011)\citenamefont
  {Huotari}, \citenamefont {Cazzaniga}, \citenamefont {Weissker}, \citenamefont
  {Pylkk\"anen}, \citenamefont {M\"uller}, \citenamefont {Reining},
  \citenamefont {Onida},\ and\ \citenamefont {Monaco}}]{huotari2011}%
  \BibitemOpen
  \bibfield  {author} {\bibinfo {author} {\bibfnamefont {S.}~\bibnamefont
  {Huotari}}, \bibinfo {author} {\bibfnamefont {M.}~\bibnamefont {Cazzaniga}},
  \bibinfo {author} {\bibfnamefont {H.-C.}\ \bibnamefont {Weissker}}, \bibinfo
  {author} {\bibfnamefont {T.}~\bibnamefont {Pylkk\"anen}}, \bibinfo {author}
  {\bibfnamefont {H.}~\bibnamefont {M\"uller}}, \bibinfo {author}
  {\bibfnamefont {L.}~\bibnamefont {Reining}}, \bibinfo {author} {\bibfnamefont
  {G.}~\bibnamefont {Onida}}, \ and\ \bibinfo {author} {\bibfnamefont
  {G.}~\bibnamefont {Monaco}},\ }\href@noop {} {\bibfield  {journal} {\bibinfo
  {journal} {Phys. Rev. B}\ }\textbf {\bibinfo {volume} {84}},\ \bibinfo
  {pages} {075108} (\bibinfo {year} {2011})}\BibitemShut {NoStop}%
\bibitem [{\citenamefont {Cazzaniga}\ \emph {et~al.}(2011)\citenamefont
  {Cazzaniga}, \citenamefont {Weissker}, \citenamefont {Huotari}, \citenamefont
  {Pylkk\"anen}, \citenamefont {Salvestrini}, \citenamefont {Monaco},
  \citenamefont {Onida},\ and\ \citenamefont {Reining}}]{cazzaniga2011}%
  \BibitemOpen
  \bibfield  {author} {\bibinfo {author} {\bibfnamefont {M.}~\bibnamefont
  {Cazzaniga}}, \bibinfo {author} {\bibfnamefont {H.-C.}\ \bibnamefont
  {Weissker}}, \bibinfo {author} {\bibfnamefont {S.}~\bibnamefont {Huotari}},
  \bibinfo {author} {\bibfnamefont {T.}~\bibnamefont {Pylkk\"anen}}, \bibinfo
  {author} {\bibfnamefont {P.}~\bibnamefont {Salvestrini}}, \bibinfo {author}
  {\bibfnamefont {G.}~\bibnamefont {Monaco}}, \bibinfo {author} {\bibfnamefont
  {G.}~\bibnamefont {Onida}}, \ and\ \bibinfo {author} {\bibfnamefont
  {L.}~\bibnamefont {Reining}},\ }\href@noop {} {\bibfield  {journal} {\bibinfo
   {journal} {Phys. Rev. B}\ }\textbf {\bibinfo {volume} {84}},\ \bibinfo
  {pages} {075109} (\bibinfo {year} {2011})}\BibitemShut {NoStop}%
\bibitem [{\citenamefont {Reining}\ \emph {et~al.}(2002)\citenamefont
  {Reining}, \citenamefont {Olevano}, \citenamefont {Rubio},\ and\
  \citenamefont {Onida}}]{lrc}%
  \BibitemOpen
  \bibfield  {author} {\bibinfo {author} {\bibfnamefont {L.}~\bibnamefont
  {Reining}}, \bibinfo {author} {\bibfnamefont {V.}~\bibnamefont {Olevano}},
  \bibinfo {author} {\bibfnamefont {A.}~\bibnamefont {Rubio}}, \ and\ \bibinfo
  {author} {\bibfnamefont {G.}~\bibnamefont {Onida}},\ }\href@noop {}
  {\bibfield  {journal} {\bibinfo  {journal} {Phys. Rev. Lett.}\ }\textbf
  {\bibinfo {volume} {88}},\ \bibinfo {pages} {066404} (\bibinfo {year}
  {2002})}\BibitemShut {NoStop}%
\bibitem [{\citenamefont {Sottile}\ \emph {et~al.}(2003)\citenamefont
  {Sottile}, \citenamefont {Olevano},\ and\ \citenamefont {Reining}}]{mt}%
  \BibitemOpen
  \bibfield  {author} {\bibinfo {author} {\bibfnamefont {F.}~\bibnamefont
  {Sottile}}, \bibinfo {author} {\bibfnamefont {V.}~\bibnamefont {Olevano}}, \
  and\ \bibinfo {author} {\bibfnamefont {L.}~\bibnamefont {Reining}},\
  }\href@noop {} {\bibfield  {journal} {\bibinfo  {journal} {Phys. Rev. Lett.}\
  }\textbf {\bibinfo {volume} {91}},\ \bibinfo {pages} {056402} (\bibinfo
  {year} {2003})}\BibitemShut {NoStop}%
\bibitem [{\citenamefont {Adragna}\ \emph {et~al.}(2003)\citenamefont
  {Adragna}, \citenamefont {{Del Sole}},\ and\ \citenamefont
  {Marini}}]{Adragna}%
  \BibitemOpen
  \bibfield  {author} {\bibinfo {author} {\bibfnamefont {G.}~\bibnamefont
  {Adragna}}, \bibinfo {author} {\bibfnamefont {R.}~\bibnamefont {{Del Sole}}},
  \ and\ \bibinfo {author} {\bibfnamefont {A.}~\bibnamefont {Marini}},\
  }\href@noop {} {\bibfield  {journal} {\bibinfo  {journal} {Phys. Rev. B}\
  }\textbf {\bibinfo {volume} {68}},\ \bibinfo {pages} {165108} (\bibinfo
  {year} {2003})}\BibitemShut {NoStop}%
\bibitem [{\citenamefont {Marini}\ \emph {et~al.}(2003)\citenamefont {Marini},
  \citenamefont {{Del Sole}},\ and\ \citenamefont {Rubio}}]{Marini}%
  \BibitemOpen
  \bibfield  {author} {\bibinfo {author} {\bibfnamefont {A.}~\bibnamefont
  {Marini}}, \bibinfo {author} {\bibfnamefont {R.}~\bibnamefont {{Del Sole}}},
  \ and\ \bibinfo {author} {\bibfnamefont {A.}~\bibnamefont {Rubio}},\
  }\href@noop {} {\bibfield  {journal} {\bibinfo  {journal} {Phys. Rev. Lett.}\
  }\textbf {\bibinfo {volume} {91}},\ \bibinfo {pages} {256402} (\bibinfo
  {year} {2003})}\BibitemShut {NoStop}%
\bibitem [{\citenamefont {Stubner}\ \emph {et~al.}(2004)\citenamefont
  {Stubner}, \citenamefont {Tokatly},\ and\ \citenamefont
  {Pankratov}}]{Tokatly}%
  \BibitemOpen
  \bibfield  {author} {\bibinfo {author} {\bibfnamefont {R.}~\bibnamefont
  {Stubner}}, \bibinfo {author} {\bibfnamefont {I.~V.}\ \bibnamefont
  {Tokatly}}, \ and\ \bibinfo {author} {\bibfnamefont {O.}~\bibnamefont
  {Pankratov}},\ }\href@noop {} {\bibfield  {journal} {\bibinfo  {journal}
  {Phys. Rev. B}\ }\textbf {\bibinfo {volume} {70}},\ \bibinfo {pages} {245119}
  (\bibinfo {year} {2004})}\BibitemShut {NoStop}%
\bibitem [{\citenamefont {{von Barth}}\ \emph {et~al.}(2005)\citenamefont {{von
  Barth}}, \citenamefont {Dahlen}, \citenamefont {van Leeuwen},\ and\
  \citenamefont {Stefanucci}}]{conskernel}%
  \BibitemOpen
  \bibfield  {author} {\bibinfo {author} {\bibfnamefont {U.}~\bibnamefont {{von
  Barth}}}, \bibinfo {author} {\bibfnamefont {N.~E.}\ \bibnamefont {Dahlen}},
  \bibinfo {author} {\bibfnamefont {R.}~\bibnamefont {van Leeuwen}}, \ and\
  \bibinfo {author} {\bibfnamefont {G.}~\bibnamefont {Stefanucci}},\
  }\href@noop {} {\bibfield  {journal} {\bibinfo  {journal} {Phys. Rev. B}\
  }\textbf {\bibinfo {volume} {72}},\ \bibinfo {pages} {235109} (\bibinfo
  {year} {2005})}\BibitemShut {NoStop}%
\bibitem [{ABI(2006)}]{ABINIT}%
  \BibitemOpen
  \href@noop {} {} (\bibinfo {year} {2006}),\ \bibinfo {note}
  {{h}ttp://www.abinit.org}\BibitemShut {NoStop}%
\bibitem [{\citenamefont {Schindlmayr}\ \emph {et~al.}(2001)\citenamefont
  {Schindlmayr}, \citenamefont {Garcia-Gonzalez},\ and\ \citenamefont
  {Godby}}]{Schindlmayr01}%
  \BibitemOpen
  \bibfield  {author} {\bibinfo {author} {\bibfnamefont {A.}~\bibnamefont
  {Schindlmayr}}, \bibinfo {author} {\bibfnamefont {P.}~\bibnamefont
  {Garcia-Gonzalez}}, \ and\ \bibinfo {author} {\bibfnamefont {R.~W.}\
  \bibnamefont {Godby}},\ }\href@noop {} {\bibfield  {journal} {\bibinfo
  {journal} {Phys. Rev. B}\ }\textbf {\bibinfo {volume} {64}},\ \bibinfo
  {pages} {235106} (\bibinfo {year} {2001})}\BibitemShut {NoStop}%
\end{thebibliography}%

\end{document}